\documentclass[a4paper,8pt]{article}

\pdfoutput=1
\usepackage{jheppub}

\usepackage{jheppub}
\usepackage{amsmath,amssymb,multirow,array}

\definecolor{green}{rgb}{0.1,0.8,0.2}

\newcommand{\half}{\frac{1}{2}}

\title{Surface transport in plasma-balls}
\author[a]{Jay Armas}
\author[b]{\hspace{-0.15cm}, Jyotirmoy Bhattacharya}
\author[c]{and Nilay Kundu}
\affiliation[a]{Physique Th{\'e}orique et Math{\'e}matique
Universit{\'e} Libre de Bruxelles and International Solvay Institutes
ULB-Campus Plaine CP231, B-1050 Brussels, Belgium}
\affiliation[b]{Centre for Particle Theory \& Department of Mathematical Sciences,
Durham University, South Road, Durham DH1 3LE, United Kingdom}
\affiliation[c]{Harish-Chandra Research Institute, Chhatnag Road, Jhunsi, Allahabad 211019, India}

\emailAdd{jarmas@ulb.ac.be, jyotirmoy.bhattacharya@durham.ac.uk, nilay.tifr@gmail.com}

\abstract{We study the surface transport properties of stationary 
localized configurations of relativistic fluids to the first two non-trivial 
orders in a derivative expansion. By demanding that these finite lumps of relativistic
fluid are described by a thermal partition function with arbitrary stationary background 
metric and gauge fields, we are able to find several constraints among surface transport coefficients. 
At leading order, besides recovering the surface thermodynamics, we obtain a generalization of the 
Young-Laplace equation for relativistic fluid surfaces, by considering a temperature dependence in the surface tension, 
which is further generalized in the context of superfluids. At the next order, for uncharged fluids in 3+1 dimensions, 
we show that besides the 3 independent bulk transport coefficients previously known, a generic localized configuration is 
characterized by 3 additional surface transport coefficients, one of which may be identified with the surface 
modulus of rigidity. 
Finally, as an application, we study the effect of temperature dependence of surface tension on some explicit examples of 
localized fluid configurations, which are dual to certain non-trivial black hole solutions via the AdS/CFT correspondence.
}
\begin{document} 

%preprint
\begin{flushright} 
% \small{preprint} 
\end{flushright}

\maketitle
\flushbottom
%
%%%%%%%%%%%%%%%%%%%%%%%%%%%%%
%
%****************************************
\section{Introduction}\label{sec:intro}
%****************************************
%
The theory of hydrodynamics provides us with a tractable effective framework to study the low-energy near-equilibrium states in any finite 
temperature system with a well behaved microscopic description. Although the description of these states in terms of the microscopic degrees of freedom
 may be very complicated, hydrodynamics allows us to describe them with a few effective degrees of freedom - the fluid 
fields. This effective theory is constructed purely on the basis of symmetries inherent to the microscopic theory, in addition to certain 
empirical assumptions like the second law of thermodynamics (see \cite{Landau:1987gn} 
\footnote{See also \cite{Son:2009tf, Bhattacharya:2011tra, Bhattacharyya:2012nq, Haehl:2015pja, Haehl:2015foa} for a more recent use and discussion 
of the second law of thermodynamics in the context of hydrodynamics.}).
The information of the underlying field theory is encoded, in a phenomenological way, in the transport coefficients that characterize the effective macroscopic description.
\footnote{These transport coefficients are often expressible in terms of correlators of symmetry currents (Kubo formulae), which may be 
evaluated directly from the microscopic quantum field theory. 
}

Although hydrodynamics is a very old and well studied subject, recently there has been a renewed interest in it, particularly after the 
discovery of its connections with black hole dynamics in the context of the AdS/CFT correspondence \cite{Bhattacharyya:2008jc}. 
These recent studies have valuably contributed to the improved understanding of the structural features of the subject and has led to 
the unraveling of new transport phenomenon \cite{Banerjee:2008th, Erdmenger:2008rm, Son:2009tf, Bhattacharya:2011eea, Herzog:2011ec, Bhattacharya:2011tra}. 
Most of these latest developments mainly focus on the description of states
in which the same phase of the fluid fills the entire spacetime, which is taken to be a non-compact pseudo-Riemanian manifold. In other words, 
the hydrodynamics that has been explored in most of these recent developments is the effective theory of a class of states, which does not 
involve any fluid surface or a phase boundary. 

In this paper we proceed to analyze the situation where the class of states described by the effective theory of hydrodynamics is extended 
to incorporate the states that include a fluid surface, which separates two phases.
We will mainly focus on finite lumps of relativistic fluids, which occupies only 
a finite subspace of an otherwise non-compact spacetime. One of the concrete 
examples of our set up is  described in \cite{Aharony:2005bm}, where metastable finite lumps of the deconfined 
phase of $\mathcal N =4$ SYM is separated from the confined phase by a phase boundary. In the large $N$ limit, such a situation can
be described by a metastable fireball of plasma-fluid separated from the vacuum by a fluid surface. \footnote{Although we have this specific set up in mind, our constructions can be straightforwardly generalized to describe the 
surface transport properties of any phase boundary.} Although we have a set up similar to \cite{Aharony:2005bm} at the back of our mind, we 
wish to clarify that in this paper we have taken a purely field theory perspective and 
we make no use of the AdS/CFT correspondence in any way. \footnote{On several occasions, in this paper, we use the word `bulk' which would always mean the bulk of the fluid in contrast to its surface. 
It should never be taken to imply the holographic dual.}

We would like to highlight the fact that the behaviour of the microscopic field theory, at or near the surface, can in general be 
quite different from that in the bulk. This difference would be captured by new surface transport coefficients
in the effective theory. Some of these new surface transport coefficients would simply encode the way in which 
the bulk transport coefficients are modified at the surface, while others would represent entirely new transport properties,
particular to the existence of the fluid surface. 

A very well known example of such a surface phenomenon, at the leading order in derivative expansion, is surface tension. In this paper, we study the surface transport coefficients, 
at the subleading order, and investigate the relations that may exists between them and the bulk transport coefficients. 
We would like to emphasize, that the surface transport coefficients carry entirely new information about the microscopics 
and  modify the fluid equations at the surface very non-trivially. For instance, knowledge of the equation of state 
\footnote{This refers to the functional dependence of the pressure of the fluid on the local temperature or energy density.} for a fluid, tells 
us nothing about how the surface tension depends on temperature. 

The study of surface transport has been carried out to some extent in the context of fluids which are confined to a thin submanifold of spacetime \cite{Armas:2013hsa, Armas:2013goa, Armas:2014rva}. This is the case in which the bulk fluid is not present or, alternatively, its pressure and higher-order transport coefficients vanish at the surface. In such situations, a systematic analysis based on an effective action approach \cite{Armas:2013hsa}, underlying symmetries and positivity of the entropy current \cite{Armas:2013goa, Armas:2014rva} have been used to constrain the form of the equations of motion up to second order in derivatives. One of the particular features inherent to the study of dynamical surfaces is that, due to possible deformations along directions transverse to it, new transport coefficients appear encoding the response of the surface to bending. One such transport coefficient is the surface modulus of rigidity \cite{Armas:2013hsa}. \footnote{In the non-relativistic context, these transport coefficients had a significant role to play in \cite{Helfrich1973, Canham197061}. See \cite{doi:10.1080/00018739700101488} for a review.} The novelty in this paper resides on the fact that the system we study is a more intricate one, 
in which both the fluid residing in the bulk and the fluid living on the surface constitute the same system.  

One of the central simplifying assumptions that we shall make in this paper is that we will only consider stationary fluid 
configuration. For the case of stationary space-filling relativistic fluids without any surface, the constitutive relations and hence 
the equations of motion could be significantly constrained from a simple physical criterion. This criterion is that we 
demand the symmetry currents, including the stress tensor, to follow from an equilibrium partition function
\cite{Banerjee:2012iz, Jensen:2012jh, Jain:2012rh, Banerjee:2012cr, Bhattacharyya:2012xi}. 

The stationary fluid configurations are considered in the presence of non-trivial background fields, 
like the metric and the gauge fields corresponding to other conserved charges. 
These background fields are considered to be slowly varying, with respect to the length scale 
associated with the radius of time-circle, in this finite temperature description. 
These slowly varying background fields serve as sources in the relativistic fluid equations.
On solving these equations, the fluid variables are expressed in terms of these background sources. 
Now, if we substitute this solution of the fluid variables, back into a putative action for stationary configurations, 
we obtain the equilibrium partition function expressed as a functional of background fields. Since, the fluid equations, 
and hence the solutions of fluids fields, are constructed in a derivative expansion, therefore the partition function 
could be expanded in a derivative expansion, in terms of these background fields and their derivatives. 

There are several advantages in considering the partition function expressed in terms of the background fields, 
(instead of the fluid variables) as the starting point of the analysis. This description is unaffected by  
any ambiguities related to choice of frames, as we move to higher order in derivatives. Also, while 
constructing the derivative expansions for the partition function, there is no need to account 
for constrains arising from lower order equation of motion, as is required while 
writing down the constitutive relations. 
\footnote{By constitutive relations, we refer to the relations expressing the symmetry currents, like the 
stress tensor etc., in terms of the fluid variables 
through a derivative expansion, based on symmetry considerations. The coefficients multiply on-shell linearly 
independent terms in this expansion.}

It is straightforward to compute the symmetry currents from the partition function by varying it with 
respect to the background fields. Then, we proceed to compare the symmetry currents so obtained, with that which 
is expressed in terms of the fluid variables through the constitutive relations. This comparison 
not only yields the expressions for the fluid variables in terms of the background fields, 
specific to the stationary configurations under consideration, but also provides non-trivial relations 
between the transport coefficients (see \S \ref{ssec:genset} for more details).
Although derived by analyzing stationary fluid configurations, these relations between transport coefficients 
are expected to hold even away from equilibrium. 
\footnote{These constraints were found to be identical to the equalities among transport coefficients that follow 
from the considerations of the second law of thermodynamics. See \cite{Bhattacharyya:2013lha, Bhattacharyya:2014bha} 
for more details on this connection.} 
In this paper, one of our principal goals would be to adopt a such a method 
to constrain transport properties at the surface of relativistic fluids. 
%
%---------------------------------------
\subsection{The general set up}\label{ssec:genset}
%---------------------------------------
%
%
%---------------------------------------
\subsubsection{Generalities of the partition function analysis}\label{sssec:genpart}
%---------------------------------------
%
Consider a relativistic fluid living in a spacetime $\mathcal N$, equipped with a time-like Killing vector, which has 
the most general stationary metric
\footnote{The discussion in this subsection is generally applicable in all dimensions, but while performing the 
analysis, particularly in \S \ref{sec:nlouncharged}, we shall specialize to four dimensions. Also we shall choose the 
Levi-Civita connection to define the covarinat derivative on $\mathcal N$.}
\begin{equation}\label{bkmet}
 ds^2 \equiv \mathcal G_{\mu \nu} dx^{\mu} dx^\nu = - e^{-2 \sigma(x)} \left(dt + a_i(x) dx^i \right)^2 + g_{ij}(x) dx^i dx^j~~.
\end{equation}
Here, the metric functions, depends only on the spatial coordinates $x$, and $\partial_t$ is our time-like Killing 
vector. Here $g_{ij}$ is the metric on spatial manifold obtained by reducing on the time-circle
\footnote{This can be done by identifying all the points on the orbits generated by the time-like Killing vector.}, 
which we shall denote by $\mathcal N_s$. 

In some of our discussions, 
we will also include a conserved global $U(1)$. The background gauge fields for this $U(1)$ take the form 
\begin{equation}\label{bkgf}
 \mathcal A = \mathcal A_0(x) dx^0 + \mathcal A_i(x) dx^i~~.
\end{equation}
Since none of the functions depend on time, all the quantities of interest, including the conserved currents, 
can be dimensionally reduced on the time-circle, whose radius we take to be $T_0$. 
It is possible to restrict to this reduced language, focusing only on $\mathcal N_s$, 
for most of the discussions relating to the partition function. 

Among the reduced quantities, time-translation invariance survives as a gauge invariance
corresponding to the Kaluza-Klein gauge field $a_i(x)$. All our constructions 
starting from the partition function, must be manifestly invariant under this Kaluza-Klein gauge
transformation. 

Since the $U(1)$ gauge fields in \eqref{bkgf} transform non-trivially under the Kaluza-Klein gauge
transformation, it is convenient to define a new set of shifted gauge fields which are manifestly invariant 
under it
\footnote{Note that our definition of $A_0$ here differs from that in \cite{Banerjee:2012iz}, in that we do not include the
shift with respect to the chemical potential, which may be thought to have been absorbed in the constant part of $\mathcal A_0$. 
% Therefore, $A_0$ directly represents the length scale associated with the chemical potential.
}
\begin{equation}\label{Adef}
 A_i =  \mathcal A_i - \mathcal A_0 a_i , ~~A_0 = \mathcal A_0~~.
\end{equation}
Thus  $\mathcal B \equiv \{ \sigma(x), a_i(x), g_{ij}(x), A_0(x), A_i(x), T_0 \}$ constitutes the set of background data in terms of which the 
partition function is to be expressed. 

Now, since we wish to describe a finite lump of relativistic fluid, we will assume that the fluid is confined to a sub-manifold 
of $\mathcal N$ of the same dimensionality as the spacetime, which we shall denote by $\mathcal M$. The fluid surface is considered to be a co-dimension 
one hypersurface. We shall denote the fluid surface by $f(x) = 0$, where $f(x)$ is taken to be 
independent of time, following our stationary assumption. In fact, $f(x)$ can be taken to be a spatial coordinate itself, which is 
positive inside the fluid and negative outside. The region inside, $f(x) = 0$ is $\mathcal M$, 
which can again be reduced on the time-circle to obtain $\mathcal M_s$. Here $\mathcal M_s$
is also a sub-manifold of $\mathcal N_s$, with a compact boundary. We furthermore assume that the boundary of $\mathcal M$ does not have boundaries itself.

The normal vector orthogonal to the fluid surface 
\begin{equation}\label{ndef}
 n_{\mu} = - \frac{\partial_\mu f(x)}{\sqrt{\partial_\nu f(x)\partial^\nu f(x)}} 
 = \left\{ 0, - \frac{\partial_i f(x)}{\sqrt{\partial_j f(x)\partial^j f(x)}} \right\}~~,
\end{equation}
is a spatial vector, with a vanishing inner product with the time-like Killing vector. 

\begin{figure}[!tbp]
  \centering
  \includegraphics[width=0.5\textwidth]{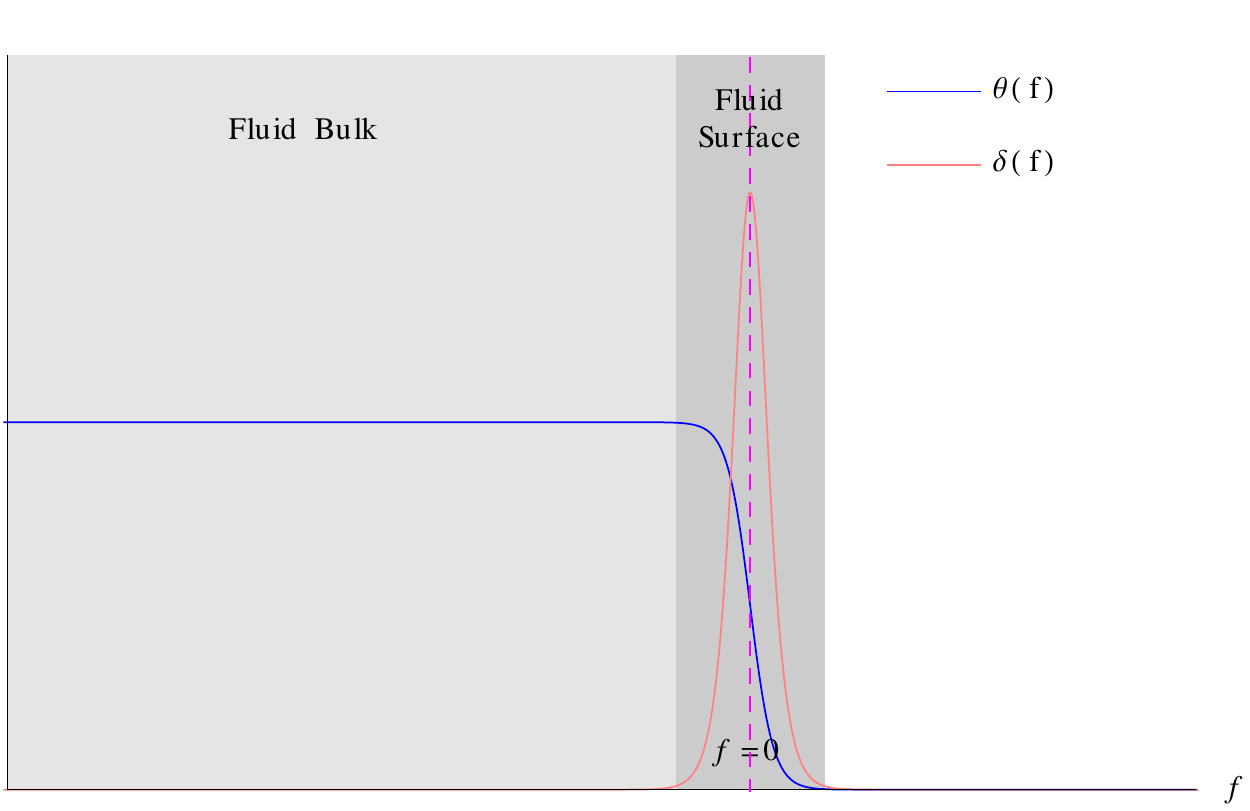}
  \caption{Schematic plots of surface distribution functionals. The dark region denotes surface while the light shaded region denotes the inside of the
  fluid lump.}\label{fig:thetadelta}
\end{figure}

The partition function of interest, after performing the trivial time integral, can be schematically written as 
\begin{equation}\label{schpf}
 \mathcal W = \int_{\mathcal N_s}  S(\mathcal B, \nabla \mathcal B \dots, \theta(f(x)), \partial \theta(f(x)) \dots)~~,
\end{equation}
where $\theta(f)$ is a distribution functional of the surface function $f(x)$ and is introduced to encode 
the variation of the fluid fields at the surface. Here, $\theta(f)$ contains all the information of the surface.
In particular, it has a dependence on the dimensionless parameter $\tilde \tau =  T_0 \tau$, with $\tau$ being the length scale associated with 
the surface thickness. All such non-universal dependence of $\theta(f)$, which are sensitive to the microscopics, 
are left implicit throughout our analysis. Realistically $\theta(f)$ is 
a distribution as shown schematically in Fig.\ref{fig:thetadelta}. The notation $\theta(f)$ is purposely used to 
indicate the fact that, in the limiting case where the parameter $\tilde \tau$ is small, 
this distribution may be well approximated by a Heaviside step function. 
\footnote{Besides the $\theta(f)$ and $\tilde \delta(f)$, the surface transport coefficients may also be dependent on 
$\tilde \tau$. Here we shall also leave that implicit (see \S \ref{ssec:sumarry} for more details).}
We will also assume that the size of the fluid configuration, i.e. the average length scale associated with 
$\mathcal M_s$, is much greater than $\tau$ as well as that associated with the temperature. 

We would like to expand $S$ in a derivative expansion. Keeping in mind the reparameterization invariance 
of the surface, this derivative expansion can be schematically performed in the following way
\begin{equation}\label{PFprim}
\begin{split}
 \mathcal W &= \int_{\mathcal N_s} \left( \theta(f)~\big( S^{\mathfrak{0}}_{(0)}(\mathcal B) + S^{\mathfrak{0}}_{(1)}(\mathcal B, \nabla \mathcal B) 
 + S^{\mathfrak{0}}_{(2)} (\mathcal B, \nabla \mathcal B, \nabla^2\mathcal B) + \dots \big)   \right. \\ 
 & \left.  \qquad + n\cdot\partial \theta(f) ~\big( S^{\mathfrak{1}}_{(0)}(\mathcal B, n) 
 + S^{\mathfrak{1}}_{(1)}(\mathcal B,\nabla \mathcal B, n , \nabla n)  
 + \dots \big)
 + \dots \right)~~  ,
 \end{split}
\end{equation}
where $S^{\mathfrak j}_{(i)}$ is a collection of terms containing $i$ number of derivatives on the background fields $\mathcal B$.
$S^{\mathfrak 0}_{(i)}$ are the terms in the bulk of the fluid and they are exactly the ones that have been considered for 
space filling fluids, in the earlier constructions of stationary fluid partition functions 
\cite{Banerjee:2012iz, Jensen:2011xb, Jain:2012rh, Banerjee:2012cr, Bhattacharyya:2012xi}. The dots 
at the end, in \eqref{PFprim}, denote terms where more than one derivatives act on $\theta(f)$. 
On all such terms, an integration by parts can be performed and they can be cast into the same form as the second term 
in \eqref{PFprim}. On performing such an integration by parts, the modified $S^{\mathfrak 1}_{(i)}$ now must 
 contain terms involving derivatives of the normal vector $n$. These new kind of terms specific to the existence 
of the surface are simply the ones involving the extrinsic curvature of the surface and its derivatives. Indeed, these are the type of terms considered in the analysis of effective actions for fluids confined to a thin surface \cite{Armas:2013hsa, Armas:2013goa, Armas:2014rva}. We may write, 
without any loss of generality
\footnote{Note that reparameterization invariance of the surface fixes the dependence on $\sqrt{\partial f \cdot \partial f}$. Therefore 
any other additional dependence on this quantity has not been considered.}
\begin{equation}\label{PFprim2}
 \mathcal W = \int_{\mathcal N_s} \theta(f) ~\big( S^{\mathfrak{0}}_{(0)} + S^{\mathfrak{0}}_{(1)} + S^{\mathfrak{0}}_{(2)} +\dots \big) + 
 \tilde \delta(f) ~\big( \tilde S^{\mathfrak{1}}_{(0)}(\mathcal B) + \tilde S^{\mathfrak{1}}_{(1)}(\mathcal B,\nabla \mathcal B,n, \nabla n)  + \dots \big)~~,
\end{equation}
where we have used 
\begin{equation} \label{derthe}
 - n \cdot \partial \theta(f) = \sqrt{\partial f \cdot \partial f} \delta(f) \equiv \tilde \delta(f)~~.
\end{equation}
Here, $\delta(f)$ denotes the derivative of the distribution $\theta(f)$. This notation is again purposely 
chosen, so that, in the limit where $\theta(f)$ approximates to Heaviside step function, $\delta(f)$ approximates to the Dirac delta function. 
Finally, if we can reliably approximate $\theta(f)$ and $\delta(f)$ to the Heaviside step function and the Dirac delta 
functions respectively, we may write \eqref{PFprim2} as 
\footnote{Here we have to use the fact, $\sqrt{\partial f \cdot \partial f} = \sqrt{\frac{\gamma}{g}}$, where $\gamma$ is the 
determinant of the induced metric on $\partial \mathcal M_s$, $\gamma_{ij} = g_{ij} - n_i n_j$.}
\begin{equation}\label{PFprim3}
 \mathcal W = \int_{\mathcal M_s} ~\big( S^{\mathfrak{0}}_{(0)} + S^{\mathfrak{0}}_{(1)} + S^{\mathfrak{0}}_{(2)}+ \dots \big) + 
 \int_{\partial \mathcal M_s} \big( \tilde S^{\mathfrak{1}}_{(0)}(\mathcal B) + \tilde S^{\mathfrak{1}}_{(1)}(\mathcal B,\nabla \mathcal B,n, \nabla n)  + \dots \big)~~.
\end{equation}
The second term in \eqref{PFprim3} is the main focus of this paper, and in particular cases, we shall provide the explicit forms 
of the surface partition function, up to the first non-trivial orders in derivatives. 

Upon variation of the partition function \eqref{PFprim2} with respect to the background metric, the stress tensor that we get 
has the form 
\begin{equation}\label{STschpf}
 T_{PF}^{\mu \nu} = T_{(0)}^{\mu \nu}\theta(f)  +T_{(1)}^{\mu \nu} \tilde \delta(f)  +T_{(2)}^{\mu \nu \rho} \partial_\rho \tilde \delta(f) + \dots~~.
\end{equation}
Note that although we were able to remove the derivatives of delta function in the partition function by an integration by parts, such derivatives are still
present in the expression for the local stress tensor. The remaining symmetry currents also have a structure similar to \eqref{STschpf}.

There is a very important and interesting role played by the function $f(x)$ in the partition function \eqref{PFprim2}. We may derive 
an equation of motion for $f(x)$ by extremizing the partition function with respect to it. We can think of this equation
as the one which determines the location of the surface. This equation of motion for $f(x)$ is identical to the particular surface fluid 
equation which follows by demanding diffeomorphism invariance in directions orthogonal to the surface. \footnote{This equation, in a limited context, is known as the Young-Laplace equation. In the later sections we will see how this Young-Laplace equation
in modified when we relax some of the assumptions made in writing its original form.}

%
%-------------------------------------------------------------------------------------------------------
\subsubsection{Fluid variables and choice of frames}\label{sssec:genpart2}
%-------------------------------------------------------------------------------------------------------
%
Now let us consider the description in terms of the original fluid variable $\{u^{\mu}, T, \dots\}$. We can write down a stress tensor 
in terms of these fluid fields and their derivatives purely based on symmetry considerations, which has the same structure as \eqref{STschpf}, namely
\begin{equation}\label{STschC}
 T_{C}^{\mu \nu} =  \tilde T_{(0)}^{\mu \nu}\theta(f) + \tilde T_{(1)}^{\mu \nu} \tilde \delta(f) 
 + \tilde T_{(2)}^{\mu \nu \rho} \partial_\rho \tilde \delta(f) + \dots ~~.
\end{equation}
The transport coefficients are the functions of scalar fluid fields, 
that multiply specific symmetry structures when $\tilde T_{(i)}^{\mu \nu}$ is further expanded in a derivative expansion. 
The number of scalar structures that goes into the partition function is, in general, much less than the allowed linearly 
independent symmetry structures which arises in the stress tensor \eqref{STschC}. Therefore comparing $T^{C}_{\mu \nu}$ with 
$T^{PF}_{\mu \nu}$ gives us very non-trivial constraints on the transport coefficients. This exercise, when executed for the 
fluid configurations with a surface, not only gives relations among surfaces transport coefficients but also relates 
them with some of the bulk transport coefficients. 

While writing \eqref{STschC}, there is a crucial issue of the choice of fluid frames. Since we do not want the fluid 
to spill out of the surface, we must require 
\begin{equation}\label{velcond}
 u^\mu n_\mu |_{f=0}= 0~~,
\end{equation}
to be true at all orders in derivative expansion. 
Also, we want the surface to be moving in the same way as the bulk of the fluid. Therefore a suitable frame choice 
should ensure that the fluid fields does not jump discontinuously at the surface. 
We should point out that one of the most popular frame choice - the Landau frame, in which $u^{\mu}$ is 
an eigen-vector of the full stress tensor, is not a suitable frame choice in this respect. This is 
because since the stress tensor has new corrections at the surface, the Landau frame choice 
introduces discontinuities in the fluid variables at the surface. Also, trying to impose the condition 
\eqref{velcond} in addition to the Landau frame condition may turn out to be more constraining than 
necessary. 

A suggestion for a suitable frame choice may be to work with a Landau frame condition only for the bulk stress tensor. \footnote{If the bulk fluid is not present, then it is possible to set the surface fluid in the Landau frame \cite{Armas:2013goa}.} That is,
we may impose 
\begin{equation}\label{framech}
 u^{\mu} \tilde T^{(0)}_{\mu \nu} = - \mathcal E u_{\nu}~~,
\end{equation}
both in the bulk and at the surface of the fluid. Here $\mathcal E$ is only the bulk energy density. This, in particular, would imply 
that the direction of energy flow at the surface of the fluid is not along the fluid flow, which is not a problem at all. 
However, although this cures one of the problems (that there would be no separate corrections to the fluid fields at the surface), 
imposing \eqref{velcond} in addition may still be over constraining, particularly at the second and higher orders in derivatives. 

This problem may be easier to visualize, if we remember the results of \cite{Banerjee:2012iz} for 3+1 dimensional uncharged fluids. In that paper, 
after comparison with the partition function, in the Landau frame, it was found that the velocity $u^\mu$, 
which is identical to the Killing vector of \eqref{bkmet} at the leading order, 
receives nontrivial corrections in terms of derivatives of the background fields at second order in the derivative expansion.
So if we now wish to impose \eqref{velcond} on that result, it would be automatically satisfied at the leading 
and first order. But at second order, it would imply non-trivial constrains involving the background field and the 
normal vector $n_\mu$, which may be too restrictive. 

Hence, the most suitable choice of frame for this problem is to choose a frame where \eqref{velcond} is a part of the frame 
fixing condition. This is possible to implement only because, at leading order the fluid velocity $u^\mu$ is proportional 
to the time-like Killing vector of \eqref{bkmet} therefore \eqref{velcond} is automatically satisfied. The higher order corrections 
can always be manipulated by a frame choice. In fact, we may foliate the bulk of the fluid $\mathcal M$, with constant $f(x)$ surfaces, 
thus extending $n_\mu$ throughout $\mathcal M$. 
Then $u^\mu n_\mu =0$ can be chosen to be part of frame fixing condition 
throughout the bulk of the fluid. 
\footnote{This choice of frame is reminiscent of the $\mu_{diss} = 0$ frame, 
in the case of superfluids, as discussed in \cite{Bhattacharya:2011eea}.} 
The remaining part of the frame choice can be implemented by imposing a condition similar to 
\eqref{framech} but projected orthogonal to $n_\mu$. We will refer to this frame as the {\it orthogonal-Landau-frame}. 
\footnote{In appendix \ref{app:Frametrans}, we provide the details for performing a transformation from the Landau-frame 
to the orthogonal-Landau-frame, in the bulk of the fluid.} While performing the partition function analysis in \S\ref{ssec:frstordpartrel}, 
we shall make this frame choice. 

There is another very interesting point of view while describing the fluid surface in terms of the fluid variables. 
We may consider two separate sets of fluids variables one in the bulk $\{u_{\mu}, T\}_{\text{blk}}$ and the other 
at the surface $\{u_{a}, T\}_{\text{sur}}$. The surface has one less fluid variable because it is one dimension less 
than the bulk. Now we can regard the surface equation of motion 
\footnote{By this we mean the part of the equation of motion proportional 
to $\tilde \delta(f)$ and its derivatives. Note that the number of fluid equations 
at the boundary is equal to the number of dimensions $\mathcal M$ rather than $\partial \mathcal M$, which is 
one higher than the number of fluid variables $\{u_{a}, T\}_{\text{sur}}$. The extra equation, may be thought of as 
the equation of motion for the function $f(x)$.}, 
as being the dynamical equations for the variables $\{u_{a}, T\}_{\text{sur}}$. 
In this equation, the bulk variables only act as sources and we can solve them to 
obtain specific solutions $\{u^{\mathfrak{s}}_{a}, T^{\mathfrak{s}}\}_{\text{sur}}$. Subsequently 
we should solve the bulk fluid equations with the boundary condition
\footnote{This boundary condition should be implemented at all orders in derivatives.}
\begin{equation} \label{eq:boundaryc}
  (u^\mu_{\text{blk}}) n_\mu |_{f=0}= 0, ~\text{and}~\{e^{\mu}_{a}u_{\mu}, T\}_{\text{blk}} |_{f=0}=\{u^{\mathfrak{s}}_{a}, T^{\mathfrak{s}}\}_{\text{sur}}~~,
\end{equation}
where $e^{\mu}_{a}$ is the projector on to the tangent space of the surface. Thus, in this point of view, the fluid equations are solved with dynamical 
\footnote{Here the word `dynamical' is to interpreted only in a restricted sense, since we only consider 
a stationary equilibrium ansatz. Some dynamics is still present in our stationary assumption, in contrast to a static one.}
boundary conditions. The dynamics of the boundary conditions are given by the surface equations of motion. In this paper, we strive to 
constrain the form of this surface equation using the framework of the equilibrium partition function.

%--------------------------------------------------------------------
\subsection{A brief summary of results}\label{ssec:sumarry}
%--------------------------------------------------------------------
%
At first, in \S \ref{sec:perfect} we consider perfect fluids in arbitrary dimensions. 
The partition function for perfect fluids with a surface can be written as
\begin{equation}\label{sum:surpartfn}
 \mathcal W = \log \mathcal Z = \int_{\mathcal{N}_s} d^3x \sqrt{g} \left( \theta(f) ~\frac{e^\sigma}{T_0} ~\mathcal P \left( T_0 e^{-\sigma}\right)
 + \tilde \delta(f) ~\frac{e^\sigma}{T_0} ~\mathcal C \left( T_0 e^{-\sigma}\right) \right)~~,
\end{equation}
where just like $\mathcal P$ can be identified with the pressure $P$ in the bulk of the perfect fluid, 
$\mathcal C$ is identified with the surface tension $\chi$. Comparison of the stress tensor 
constructed through symmetry arguments, with that following from \eqref{sum:surpartfn} yields the expected 
surface thermodynamics. 
The component of the stress tensor conservation equation normal to the surface, 
at the surface, reads
\begin{equation}\label{sum:genLY}
 P (T) |_{f=0} =\chi K + T \chi_S ~n_{\mu} \mathfrak{a}^{\mu} |_{f=0} ~~, 
\end{equation}
where $K$ is the extrinsic curvature of the surface and $\mathfrak{a}_\mu$ is 
the fluid acceleration. This is a modified version of the Young-Laplace equation where the term proportional to the 
acceleration is new compared to its original form. This additional term is non-zero only 
if the surface entropy $\chi_S$ is non-zero, i.e. , if there is a non-trivial
temperature dependence of surface tension. 

This term has a very simple physical interpretation. If the surface entropy 
is non-zero, it implies that there are non-trivial degrees of freedom localized 
at the surface. Then the additional term accounts for the centripetal acceleration
of these degrees of freedom in the force balance equation that \eqref{sum:genLY}
represents. 

Since the acceleration term in \eqref{sum:genLY} has not been widely considered in 
the literature before, we analyzed the consequence of this term  on some 
simple fluid configurations in \S \ref{sec:config}. For this purpose, 
as a sample system, we choose to revisit the localized configurations of deconfined 
plasma of large $N$, strongly coupled $\mathcal N = 4$ Yang-Mills theory, compatified down to $2+1$ dimensions 
on a Scherk-Schwarz circle, that were constructed in \cite{Lahiri:2007ae}. These 
configurations are dual to exotic black holes in Scherk-Schwarz compactified $AdS_5$. 

However, due to the unavailability of the exact dependence of surface tension on temperature for this system, 
 the surface tension was taken to be a constant 
\footnote{The value of surface tension at the phase-transition temperature which was previously computed in 
\cite{Aharony:2005bm} was used for this constant value.}, in the analysis of \cite{Lahiri:2007ae}. 
In \S \ref{sec:config}, we suitably parameterized this ignorance and studied the 
change in the phase diagram for the configurations, as we varied this parameter. We found that turning on 
this parameter introduced an upper bound on the surface velocity. This arises from the fact that 
the surface temperature dips below the phase transition temperature, when the bound is overshot. This 
results in the termination of the phase curve at a specific point, see fig. \ref{fig:pdgm}. This has important consequences for the existence of a phase transition between the ball and ring configurations. 
We find that the phase transition may not exist for large values of this parameter
\footnote{More specifically, if $\chi \sim T^{-\zeta}$, then we found for $\zeta \gtrsim 1/3$ the phase transition would cease to exist
between the ball and the ring.}. 

Moving on to the case of finite lumps of superfluids, at zeroth order in derivatives, $\mathcal P$ and $\mathcal C$ 
in the partition function \eqref{sum:surpartfn}, now would also depend on $A_0$ and the norm of the superfluid velocity 
$\xi_{\mu}$ \cite{Bhattacharyya:2012xi}. We find that \eqref{sum:genLY} is further modified in the case of superfluids to become 
\begin{equation}
  P (T) |_{f=0} = \chi K+\left( \chi_E - \chi \right) ~n_{\mu} \mathfrak{a}^{\mu} |_{f=0} + \lambda ~ n^\mu ~\xi^\nu ~\nabla_\nu \xi_\mu |_{f=0}~~. 
\end{equation}

In \S \ref{sec:nlouncharged} we consider the case of uncharged fluids in 3+1 dimensions, where the first corrections to the 
perfect fluid partition function occurs at second order in the bulk and at first order on the surface. The full partition function 
upto this order, including the parity odd sector, takes the form
\begin{equation}\label{sum:surpartsecord}
\begin{split}
 & \mathcal W = \log \mathcal Z =  \int_{\mathcal{M}_s} d^3x \sqrt{g} \left(\frac{e^\sigma}{T_0} ~\mathcal P \left( T_0 e^{-\sigma}\right)  
%  \\ &
 -\frac{1}{2} \bigg[ P_1(\sigma) R + T_0^2  P_2(\sigma) f_{ij}f^{ij} 
 + P_3(\sigma)( \partial \sigma)^2 
%  + P_4(\sigma) \nabla^2 \sigma
 \bigg] \right) \\
&+ \int_{\partial {\mathcal{M}_s}} ~d^2x \sqrt{\gamma} \frac{e^\sigma}{T_0}  \left(  ~\mathcal C \left( T_0 e^{-\sigma}\right) 
 + \mathcal{B}_1 \left( T_0 e^{-\sigma}\right) n^i~\partial_i \sigma 
 + \mathcal{B}_2\left(T_0 e^{-\sigma}\right) \epsilon^{ijk} n_i f_{jk} + \mathcal B_{3}\left(T_0 e^{-\sigma}\right) \mathcal K
 \right) \bigg|_{f=0}~~.
 \end{split}
\end{equation}
Here $P_i$ are the three independent coefficients that were considered in \cite{Banerjee:2012iz}, while $\mathcal B_i$ are 
the new three surface transport coefficients. The terms proportional to $\mathcal B_1$ and $\mathcal B_2$ 
can also be viewed as bulk total derivative terms, while the term $\mathcal B_3$, 
a term which was studied in \cite{Armas:2013hsa}, eventually contributes to the 
modulus of rigidity. 

As pointed out before, the surface transport coefficients in \eqref{sum:surpartsecord} may also 
depend on $\tilde \tau$ (the dimensionless ratio of surface thickness and $T_0$), which we leave implicit here. 
If any particular limit is taken on this parameter $\tilde \tau$, it may directly influence 
the surface transport coefficients in \eqref{sum:surpartsecord}, particularly $\mathcal{B}_3$.

For the stress tensor $T_{C}^{\mu \nu}$, which follows from symmetry considerations, there are 31 surface terms 
that can be written down, which are linearly independent for stationary configurations. Now, taking 
into consideration the fact that $P_i$ correspond to three independent transport coefficient 
in the bulk fluid, we are able to derive 28 relations between the 31 surface transport coefficients 
and the 3 independent bulk transport coefficients, as it has been explicated in \S \ref{sec:nlouncharged}. 
\section{Perfect fluids}\label{sec:perfect}
%****************************************
%
In this section we will study fluids at zeroth order in derivatives. At this order, the only surface effect is encoded in the
surface tension, which is extremely well studied. However, it is very instructive to re-derive the known physics 
in the language of partition functions described in \S \ref{sec:intro}. We will also get the occasions to discuss 
a few effects related to the temperature dependence of surface tension and surface tension in superfluids which has 
not been widely discussed in the literature. 
% 
%-----------------------------------------------------------------------------------------------------------------
\subsection{Ordinary uncharged perfect fluids in arbitrary dimensions}\label{ssec:perfctfl}
%-----------------------------------------------------------------------------------------------------------------
%  
At first, let us briefly review the partition function for space filling ordinary perfect fluids as discussed in \cite{Banerjee:2012iz}.
The partition function in terms of the metric sources can be written as, 
\begin{equation}\label{pfpartfn}
 \mathcal W = \log \mathcal Z = \int d^3x \sqrt{g} ~\frac{e^\sigma}{T_0} ~\mathcal P \left( T_0 e^{-\sigma}\right)~~.
\end{equation}
The functional form of $\mathcal P$ is to be determined from microscopics. 
Let us now evaluate the stress 
tensor from the above partition function by using \cite{Banerjee:2012iz}
\begin{equation}\label{stdef}
 T_{00} = - \frac{T_0 e^{2 \sigma}}{\sqrt {-\cal G}} \frac{\delta \mathcal W}{\delta \sigma} ~~,~~
 T_{0}^{i} = \frac{T_0}{\sqrt {-\cal G}} \frac{\delta \mathcal W}{\delta a_i} ~~,~~
 T^{ij} = - \frac{2 T_0}{\sqrt {-\cal G}} g^{il} g^{jm} \frac{\delta \mathcal W}{\delta g^{lm}}~~.
\end{equation}
Evaluating these formulae explicitly for \eqref{pfpartfn} we get 
\begin{equation}\label{perfSTpf}
  T_{00} = e^{2 \sigma} \left(\mathcal P -  T \frac{\partial \mathcal P }{\partial  T}\right)~~,
  ~T_{0}^{i} = 0~ ,~T^{ij} = \mathcal P g^{ij}~~,
\end{equation} 
where $T =T_0  ~e^{-\sigma}$.
By comparing \eqref{perfSTpf} with the zeroth order form of the stress tensor that follows from symmetry considerations
\begin{equation} \label{pfstress}
  T_{\mu \nu} = \left(\mathcal E (T) + P(T) \right) u_{\mu} u_{\nu} + P(T) \mathcal{G}_{\mu \nu}+ \dots~~,
\end{equation}
we get 
\begin{equation}\label{pfiden}
 P = \mathcal P~,~ \mathcal E = - \mathcal P +  T \frac{\partial \mathcal P }{\partial  T}~~,
\end{equation}
while the fluids fields are found to be
\begin{equation} \label{flvarsol0}
 u^{\mu} = e^{- \sigma} \{1,0,\dots, 0\} + \dots~~, ~~T = T_{0} e^{-\sigma}+ \dots ~~.
\end{equation}
Note that \eqref{pfiden} is identical to the condition on pressure and energy density that follows from thermodynamics. 
In this way, we are able to derive the thermodynamic properties of the fluid by comparison with the partition function. In some sense, 
pressure and energy density can be thought of as zeroth order transport coefficients, which are related by thermodynamic relations which follow from the partition function analysis. 

Following the above procedure, we wish to write down a partition function for 
perfect fluids in equilibrium, confined within a surface (which itself is dynamically 
determined by minimization of the free energy). Respecting the principles of KK-gauge invariance 
for writing down the partition function and reparameterization invariance 
of the surface, the partition function is given in terms of two unknown functions
\footnote{
Note that, as described in \S \ref{sec:intro}, under suitable assumptions on $\theta(f)$ and $\delta(f)$, \eqref{surpartfn} may also be written as,
\begin{equation}\label{surpartfnnew}
 \mathcal W = \log \mathcal Z = \int_{\mathcal{M}_s} d^3x \sqrt{g}\frac{e^\sigma}{T_0} ~\mathcal P \left( T_0 e^{-\sigma}\right)
 + \int_{\partial {\mathcal{M}_s}} ~d^2x \sqrt{\gamma} \frac{e^\sigma}{T_0} ~\mathcal C \left( T_0 e^{-\sigma}\right) \bigg|_{f=0}~~.
\end{equation}
}
\begin{equation}\label{surpartfn}
 \mathcal W = \log \mathcal Z = \int_{\mathcal{N}_s} d^3x \sqrt{g} \left( \theta(f) ~\frac{e^\sigma}{T_0} ~\mathcal P \left( T_0 e^{-\sigma}\right)
 + \tilde \delta(f) ~\frac{e^\sigma}{T_0} ~\mathcal C \left( T_0 e^{-\sigma}\right) \right)~~.
\end{equation}
In order to obtain the stress tensor, we have to vary the partition function \eqref{surpartfn} with respect to the background metric fields 
\footnote{The functional $\mathcal W$ is to be taken to be a functional of the metric functions and the function $f$,
defining the surface. All these functions are independent functions of the coordinates and must be treated independently. }.

In fact, using \eqref{stdef}, explicitly we find
\begin{equation}\label{pfstsurften}
\begin{split}
 T_{00} &= e^{2 \sigma} \left(\mathcal P -  T \frac{\partial \mathcal P }{\partial T}\right)\theta(f)
 +\tilde \delta(f)e^{2 \sigma} \left(\mathcal C -T \frac{\partial \mathcal C }{\partial  T}\right)~~,\\
  T_{0}^{i} &= 0~ ~,\\
  T^{ij} &= \mathcal P g^{ij}\theta(f)+\tilde \delta(f)\mathcal C \gamma^{ij}~~.
  \end{split}
\end{equation} 
Now we have to compare \eqref{pfstsurften} with the stress tensor that may be written from symmetry arguments \eqref{STschC} to this particular order, namely,
\begin{equation}\label{fullstress}
 T_{\mu \nu} = T_{\mu \nu}^{(0)}\theta(f) + T_{\mu \nu}^{(1)}\tilde \delta(f)~~.
\end{equation} 
The bulk stress tensor components $T_{\mu \nu}^{(0)}$ are given by \eqref{pfstress}, while the components of the surface stress 
tensor also have a similar form
\begin{equation}\label{surstress}
 T_{\mu \nu}^{(1)} =  \chi_E (T) ~ u_{\mu} u_{\nu} - \chi(T) ~\mathcal P_{\mu \nu}   + \dots~~.
\end{equation}
Here, $\mathcal {P}_{\mu \nu} = \mathcal{G}_{\mu\nu} + u_{\mu} u_{\nu} - n_{\mu} n_{\nu} $ is the projector orthogonal 
to both the velocity vector and the normal to the surface. In \eqref{surstress} we also introduced $\chi_E$ and $\chi$, which are, 
respectively, the surface energy and the surface `pressure', also known as surface tension.

Comparing \eqref{pfstsurften} with \eqref{surstress}, we can therefore identify
\begin{equation} \label{idchi}
\chi=- \mathcal C~~,~~\chi_{E}=-\mathcal{C}+ T\frac{\partial\mathcal{C}}{\partial T}~~,
\end{equation}
while the fluids fields are again given by \eqref{flvarsol0}.\footnote{Note that with this the continuity of the fluids fields, as we move from the bulk to the boundary is maintained. Also, 
\eqref{flvarsol0} implies $u \cdot n = 0$, is automatically satisfied, at the order of perfect fluids.} This identification \eqref{idchi} had also been done in \cite{Armas:2013hsa}. 
Just like the bulk perfect fluid, \eqref{idchi} implies the thermodynamic identity
\begin{equation}\label{surtherm}
 \chi_E = \chi + T \chi_S~~,
\end{equation}
where $\chi_S=-\partial \chi / \partial T$ is the surface entropy. Therefore, if the surface tension depends non-trivially 
on the fluid temperature, it means that the surface entropy is non-zero and hence that there are active degrees of freedom living
on the surface of the fluid. When the surface entropy vanishes (that is the surface tension is constant), the surface tension 
is equal to the negative of surface energy. 

The conservation of the stress tensor \eqref{fullstress}, implies 
\begin{equation}\label{stconvbs}
 \nabla_{\mu} T^{\mu \nu} = \theta(f) \nabla_{\mu} T^{\mu \nu}_{(0)} 
 + \delta(f) \left(- n_\mu T^{\mu \nu}_{(0)} + \tilde \nabla_{\mu} T^{\mu \nu}_{(1)} \right) + \dots~~,
\end{equation}
where we have defined 
$\tilde \nabla_\mu \left( \dots \right) = 1/\sqrt{\partial f \cdot \partial f } ~~\nabla_\mu \left( \sqrt{\partial f \cdot \partial f } \dots  \right)$.
In the bulk \eqref{stconvbs} will give rise to the usual equation bulk 
conservation equation $\nabla_\mu T^{\mu\nu}_{(0)}$, while at the surface, it gives rise to the condition
\begin{equation}\label{bdystcon}
-n_\mu T^{\mu \nu}_{(0)} |_{f=0} + \tilde \nabla_{\mu} T^{\mu \nu}_{(1)} = 0~~.
\end{equation}
This equation is in fact a Carter equation with a force term \cite{Carter:2000wv}, and is extensively used in the context of (mem)-brane hydrodynamics \cite{Emparan:2009at, Armas:2013hsa, Armas:2013goa, Armas:2014rva}. As with any Carter equation, \eqref{bdystcon} gives rise to two physically different sets of equations, obtained by projecting both orthogonally and tangentially
to the fluid surface. Before doing so, let us note that
\begin{equation}
n_\mu T^{\mu \nu}_{(0)} = n^\nu P(T) |_{f=0}~~,
\end{equation}
that is, for a perfect fluid, the bulk contribution to \eqref{bdystcon} is only present 
in the normal component of \eqref{bdystcon}. Thus, if we project \eqref{bdystcon}
along the fluid surface with the projector $e^a_{\nu}$ such that $n^\nu e^a_{\nu}=0$, we obtain 
\begin{equation} \label{tangential}
 e^a_{\nu} \tilde \nabla_\mu T^{\mu \nu}_{(1)} = 0~~.
\end{equation}
Here the index $a=0,1,2$ labels the directions along the surface. Equation \eqref{tangential} expresses the conservation of the surface stress tensor along the surface. Note that 
if we consider higher derivative corrections, this equation will also receive a contribution 
from the bulk stress tensor, which would signify energy and momentum transport from the 
bulk to the surface. For the perfect fluid, however, such transport does not take place. 

The component of \eqref{bdystcon} normal to the surface, describing the elastic degrees of freedom of the surface, is more interesting and provides 
us with the condition that determines the position of the surface. For 
perfect fluids, it reduces to 
\begin{equation}\label{surconn}
-P (T) |_{f=0} + n_{\nu} \tilde \nabla_{\mu} T^{\mu \nu}_{(1)} = 0~~.
\end{equation}
Now, given that the surface stress tensor \eqref{surstress} is orthogonal to the normal vector $n_{\mu}$,
we can rewrite the second term in \eqref{surconn} in the following way
\begin{equation}\label{surconnK}
P (T) |_{f=0} + T^{\mu \nu}_{(1)} K_{\mu \nu} = 0~~, 
\end{equation}
where $K_{\mu \nu}$ is the extrinsic curvature of the surface. We can easily check that both 
\eqref{surconn} and \eqref{surconnK} reduces to 
\footnote{Note that $u_{\mu} u_{\nu} K^{\mu \nu} = -n_{\mu} \mathfrak{a}^{\mu}$, since $n^{\mu} u_\mu =0$.}
\begin{equation}\label{genLY}
 P (T) |_{f=0} =\chi K+ \left( \chi_E - \chi \right) ~n_{\mu} \mathfrak{a}^{\mu} |_{f=0} ~~,
\end{equation}
where $\mathfrak{a}_\mu = u^{\nu} \nabla_\nu  u_{\mu}$ is the fluid acceleration, 
and $K=\mathcal{G}^{\mu\nu}K_{\mu\nu}$ is the mean extrinsic curvature. If there are no active 
degrees of freedom in the boundary and the surface tension is constant implying that 
$\chi_E = \chi$, then \eqref{genLY} immediately reduces to 
\begin{equation}\label{LY}
 P (T) |_{f=0} =  \chi  K~~,
\end{equation}
which is the Young-Laplace equation.
% % % % % % % \footnote{Note that the negative sign in this equation is 
% % % % % % % a consequence of the choice of sign of $\chi$ in \eqref{surstress}. In \cite{Bhattacharya:2009gm} 
% % % % % % % the sign of $\chi$ was chosen to be the opposite.} 
Thus \eqref{genLY} is a generalization 
of the Young-Laplace equation, when there are non-trivial degrees 
of freedom living on the surface of the fluid. Such generalization had not been previously 
considered in the works of \cite{Lahiri:2007ae, Caldarelli:2008mv, Bhattacharya:2009gm}.
In \S\ref{sec:config}, we shall examine the consequence of the presence of this additional term in \eqref{genLY} for 
some simple fluid configurations. 

It is noteworthy that \eqref{genLY} can also be directly obtained from the partition function \eqref{surpartfn}.
In terms of the partition function, this is simply given by the extremization of the 
partition function with respect to the surface function $f$, 
which is the equation of motion $f$. This is intuitively 
expected since the location of the fluid surface is obtained 
by the minimization of the free energy. If we vary \eqref{surpartfn}
with respect to $f$, at the leading order we find
\begin{equation}\label{PFLYeq}
 \delta (f) \left( \mathcal P + \mathcal C(T_0 e^{-\sigma}) \mathcal K 
 +\left(  \mathcal C(T_0 e^{-\sigma}) - \mathcal T ~\frac{\partial\mathcal C(T_0 e^{-\sigma})}{\partial\mathcal{T}} \right) ~n^{i}\partial_{i} \sigma
 \right) = 0 ~~,
\end{equation}
where $\mathcal K = g^{ij} \bar \nabla_i n_j$, with $\bar \nabla$ being the spatial covariant derivative defined with respect to the reduced metric
\footnote{$\mathcal K$ is related to the full extrinsic curvature by $K = \mathcal K + n \cdot \partial \sigma$.}.
Given the thermodynamical relations \eqref{pfiden}, \eqref{idchi} and remembering that in terms of the background fields the fluid 
acceleration is given by $\mathfrak{a}^i= g^{ij} \partial_j \sigma$ \cite{Banerjee:2012iz}, \eqref{genLY} reduces to \eqref{PFLYeq}.

%********************************************************************************
\subsection{Zeroth order superfluids with a surface}\label{sec:superfl}
%********************************************************************************

For the case of superfluids, the zeroth order stress tensor is modified in order to include the superfluid velocity $\xi_\mu$ 
\cite{Landau:1941,Tisza:1947zz}. 
The bulk stress tensor has the form
\begin{equation}
 T_{\mu \nu}^{(0)} = (\epsilon + P) u_{\mu} u_{\nu} + P \mathcal G_{\mu \nu} + \lambda \xi_{\mu} \xi_\nu ~~,
\end{equation}
and it is accompanied by a conserved current 
\begin{equation}
 J_{\mu}^{(0)} = q u_\mu - \lambda \xi_{\mu} ~~.
\end{equation}
Just like in the case of ordinary fluids, in the presence of a surface there are surface stress tensor and current contributions, which read
\begin{equation}
 \begin{split}
  T^{(1)}_{a b} &=  (\epsilon + P) u_{a} u_{b} 
  + P \mathcal H_{a b} + \lambda \xi_{a} \xi_b ~~,~~
  J^{(1)}_{a} = q ~u_a - \lambda ~\xi_{a} ~~.
 \end{split}
\end{equation}
It can be explicitly checked that this form of the boundary stress tensor and current follows from the 
following partition function
\begin{equation}
 \mathcal W = \log \mathcal Z = \int_{\mathcal{M}} d^3x \sqrt{g} \frac{e^{\sigma}}{T_0} ~\mathcal P \left(T_0 e^{-\sigma}, A_0, \xi \right) 
 + \int_{\partial {\mathcal{M}}} ~d^2x \sqrt{\gamma}  \frac{e^{\sigma}}{T_0}  ~\mathcal C \left( T_0 e^{-\sigma}, A_0, \xi \right) \bigg|_{f=0}~~,
\end{equation}
where $\xi$ is the norm of the superfluid velocity. The bulk term of this partition function was first derived in \cite{Bhattacharyya:2012xi}. In the presence of the surface we must also assume 
that the superfluid velocity and the normal to the surface are mutually orthogonal $n^{\mu} \xi_\mu |_{f=0}= 0$. 
As in the previous sections, the stress tensor is conserved and the normal component of the conservation of 
the boundary stress tensor gives the generalized Young-Laplace equation. To leading order, the bulk and 
surface currents are conserved separately, with no current flowing from the bulk to the surface. 

The generalized Laplace-Young equation \eqref{genLY}, is further modified in the case of superfluids. It takes 
the form
\begin{equation}\label{superYL}
  P (T) |_{f=0} = \chi K+\left( \chi_E - \chi \right) ~n_{\mu} \mathfrak{a}^{\mu} |_{f=0} + \lambda ~ n^\mu ~\xi^\nu ~\nabla_\nu \xi_\mu |_{f=0}~~. 
\end{equation}
Note that the new term is present even if there is no temperature dependence in the surface tension, as long as the goldstone boson 
also constitutes an active degree of freedom on the surface. Also this modified equation is also applicable 
to the case when there is an emergent goldstone boson only at the 
surface of a fluid, a situation which is reminiscent of topological insulators in the 
context of fluids.

%********************************************************************************
\section{Next to leading order corrections for uncharged fluids}\label{sec:nlouncharged}
%********************************************************************************

In this section we shall consider the next to leading order corrections for uncharged fluids with a surface. 
The principal goal of this section 
is to demonstrate that there are only three new equilibrium transport coefficients on the surface of the 
uncharged fluid, at the next to leading order, two of which are parity even and the 
other one being parity odd. Two of these new boundary terms in the partition function,
also precisely coincide with two possible bulk total derivative terms. Here, we 
work out the interplay between these new surface transport coefficients 
and the bulk second order transport coefficients. 

%********************************************************************************
\subsection{Partition function at next to leading order}\label{ssec:frstordpart}
%********************************************************************************
%
In order to write down the first corrections to the partition function \eqref{surpartfn},  
we need to write down all KK-gauge invariant scalar terms at higher order in derivatives. 
As it was observed in \cite{Banerjee:2012iz}, the bulk of the fluid does not receive any corrections 
at first order. In other words, there are no KK-gauge invariant scalar bulk terms at first order, which can be 
written in terms of the sources of an uncharged fluid. 

However, at the surface of the fluid, there is an additional geometrical structure, which is 
the vector normal to the fluid surface. This allows us to write down two possible parity even scalar terms 
which may constitute the partition function. These are
\begin{equation}\label{indtermsca1ord}
 \mathcal K ~~, ~~ n_i ~\partial^i \sigma~~.
\end{equation}
Here $\mathcal K$ is the trace of the extrinsic curvature reduced along the time direction

Now for parity odd terms, there is only one possible parity odd scalar 
% % that can be written 
% % down at the surface, which is first order in derivatives
\begin{equation}
 \epsilon^{ijk} n_i f_{jk}~~,
\end{equation}
which must be taken into account while writing down the partition function. For space-filling fluids, it was 
not possible to construct any parity odd term in the partition function, at the second order \cite{Banerjee:2012iz}. 
Therefore the existence of this term suggests that even uncharged fluids may have parity odd transport when 
surface effects are considered. Since, the fluid surface is co-dimension one, this term is particularly 
reminiscent of a parity odd transport that can exist in $2+1$ dimensions \cite{Banerjee:2012iz, Jensen:2012jh}.

Including these new surface terms, the partition function takes the following form
\begin{equation}\label{surpartfrstord}
\begin{split}
 &\mathcal W = \log \mathcal Z =  \int_{\mathcal{M}_s} d^3x \sqrt{g} \left(\frac{e^\sigma}{T_0} ~\mathcal P \left( T_0 e^{-\sigma}\right) \right)\\
 &+ \int_{\partial {\mathcal{M}_s}} ~d^2x \sqrt{\gamma} \frac{e^\sigma}{T_0}  \left(  ~\mathcal C \left( T_0 e^{-\sigma}\right) 
 + \mathcal{B}_1 \left( T_0 e^{-\sigma}\right) n^i~\partial_i \sigma 
 + \mathcal{B}_2\left(T_0 e^{-\sigma}\right) \epsilon^{ijk} n_i f_{jk} + \mathcal B_{3}\left(T_0 e^{-\sigma}\right) \mathcal K
 \right) \bigg|_{f=0}~.
 \end{split}
\end{equation}
It is important to note that if we include these first order surface terms in the partition function then we must also include 
the second order bulk terms for consistency. For instance, the surface stress tensor following from the partition function in 
\eqref{surpartfrstord} may get contributions from total derivative terms at second order in the bulk. 
In fact, we must point out that two of the new terms that we have added in \eqref{surpartfrstord}
may also be written as a bulk total derivative terms at second order.

Thus, including the bulk second order terms, which was written down in \cite{Banerjee:2012iz}
\footnote{We entirely follow the notation and conventions of \cite{Banerjee:2012iz} for the second order terms.}, we have 
\begin{equation}\label{surpartsecord}
\begin{split}
 & \mathcal W = \log \mathcal Z =  \int_{\mathcal{M}_s} d^3x \sqrt{g} \left(\frac{e^\sigma}{T_0} ~\mathcal P \left( T_0 e^{-\sigma}\right) 
%  \\ &
 -\frac{1}{2} \bigg[ P_1(\sigma) R + T_0^2  P_2(\sigma) f_{ij}f^{ij} 
 + P_3(\sigma)( \partial \sigma)^2 
%  + P_4(\sigma) \nabla^2 \sigma
 \bigg] \right) \\
&+ \int_{\partial {\mathcal{M}_s}} ~d^2x \sqrt{\gamma} \frac{e^\sigma}{T_0}  \left(  ~\mathcal C \left( T_0 e^{-\sigma}\right) 
 + \mathcal{B}_1 \left( T_0 e^{-\sigma}\right) n^i~\partial_i \sigma 
 + \mathcal{B}_2\left(T_0 e^{-\sigma}\right) \epsilon^{ijk} n_i f_{jk} + \mathcal B_{3}\left(T_0 e^{-\sigma}\right) \mathcal K
 \right) \bigg|_{f=0}~,
 \end{split}
\end{equation}
where $P_i$ denote the three independent transport coefficients at second order for a fluid without surfaces.

Now, as pointed out before we can write the new surface terms as a bulk term in the following way
\begin{equation}\label{surpartsecord2}
\begin{split}
 \mathcal W = \log \mathcal Z = & \int_{\mathcal{M}_s} d^3x \sqrt{g} \left(\frac{e^\sigma}{T_0} ~\mathcal P \left( T_0 e^{-\sigma}\right) \right.
 \\ & \left.
 -\frac{1}{2} \bigg[ P_1(\sigma) R + T_0^2  P_2(\sigma) f_{ij}f^{ij} 
 + P_3(\sigma)( \partial \sigma)^2 
+  \nabla^2 P_4(\sigma) + \epsilon^{ijk} \left( \partial_i P_5 \right)f_{jk}
 \bigg] \right) \\
 & + \int_{\partial {\mathcal{M}_s}} ~d^2x \sqrt{\gamma} \left( \frac{e^\sigma}{T_0} ~\mathcal C \left( T_0 e^{-\sigma} \right)  
 + P_6(\sigma) \mathcal K
%  + \frac{e^\sigma}{T_0} \vartheta \left( T_0 e^{-\sigma}\right) n^i~\partial_i \sigma 
 \right) \bigg|_{f=0}~,
 \end{split}
\end{equation}
where we have defined
\begin{equation}\label{vartP}
 \begin{split} 
 P_4'(\sigma) &= - \frac{2 e^{\sigma}}{T_0} \mathcal B_1 \left( T_0 e^{- \sigma}\right)~~, ~~
 P_5(\sigma) = - \frac{2 e^{\sigma}}{T_0} \mathcal B_2 \left( T_0 e^{- \sigma}\right)~~, ~~
 P_6(\sigma) = \frac{e^{\sigma}}{T_0} \mathcal B_3 \left( T_0 e^{- \sigma}\right)~~.
%   P_4(\sigma) &= -2 \frac{e^{\sigma}}{T_0} \vartheta \left( T_0 e^{- \sigma}\right)\\
%   \tilde P_3 (\sigma) &= P_3 - 2 \frac{e^{\sigma}}{T_0} \vartheta \left( T_0 e^{- \sigma}\right) + 2 \vartheta' \left( T_0 e^{- \sigma}\right)
 \end{split}
\end{equation}
Now, as it was shown in \cite{Banerjee:2012iz}, $P_1, P_2$ and $P_3$ were determined in terms of the 
bulk second order equilibrium coefficients. Also eliminating the $P_1, P_2$ and $P_3$ from those relations, 
gave rise to 5 relations among the eight possible 
second order equilibrium transport coefficients.

It is clear that the terms proportional to $P_4$ and $P_5$ (or any total derivative term) will not enter the bulk stress tensor
\footnote{This is because, a bulk total derivative can always be written as 
\begin{equation}
 \int_{\mathcal M_s} d^3x \sqrt{g} ~\nabla_i V^i = \int_{\mathcal M_s} d^3x \partial_i \left( \sqrt{g} ~V^i \right)~~,
\end{equation}
for any $V_i$. Since the variation of such terms, with respect to the metric, always lies within the derivative, 
hence such terms can only contribute to the surface stress tensor and never to the bulk stress tensor.
}.
But they definitely contribute non-trivially to the surface stress tensor. It is important to note that 
both the form of the partition function \eqref{surpartsecord} and \eqref{surpartsecord2} are equivalent and describe the 
same system. Hence, everything physical that is evaluated from them, such as the surface stress tensor or the 
Young-Laplace equation, must be identical.

%********************************************************************************
\subsection{Corrections to the stress tensor}\label{ssec:frstordpartSTYL}
%********************************************************************************
%
Once we have written down the partition function \eqref{surpartsecord2}, it is immediate to evaluate the 
stress tensor by varying with respect to the background fields using \eqref{stdef}. It is convenient to split the bulk and surface contributions, up to second order, in the followind way
\begin{equation}
T^{\mu\nu}_{\text{blk}}=T^{\mu\nu}_{(0)}\theta(f)~~,~~T^{\mu\nu}_{\text{sur}}=T^{\mu\nu}_{(1)}\tilde \delta(f)+T^{\mu\nu\rho}_{(2)}\partial_\rho\tilde \delta(f)~~.
\end{equation}
The bulk stress tensor remains the same as that computed in \cite{Banerjee:2012iz} and involves only the the coefficients $P_1, P_2$ and $P_3$ while the surface contribution is obtained as terms proportional to $\tilde \delta(f)$ and its derivatives
% % \footnote{\JB{We may want to kill this footnote later}: While performing the 
% % second integration by parts for certain terms, in order to obtain the surface stress tensor, the surface terms 
% % has been dropped. This is because those surface terms are to be evaluated on the boundary of $\mathcal N$ 
% % and not $\mathcal M$, which we consider to be vanishing due to the absence of any fluid over there. This statement 
% % may also be interpreted as the fact the boundary terms at the surface vanish.}
\footnote{Here we have kept the term proportional to $\nabla_i n_j-\nabla_j n_i$. This term depends on how $n_i$ is extended 
away from the fluid surface. We may choose to perform this extension so that this anti-symmetric derivative of $n_i$ is zero. However, 
we perform our analysis without such an assumption so that, if our equations is applied in some 
generalized circumstance where a more natural extension of $n_i$ away from the surface demands this 
term to be non-zero.}
\begin{equation}\label{STsurPF}
 \begin{split}
  {T_{\text{sur}}}^{lk} &= g^{li} g^{kj} \Bigg[ \tilde \delta(f) T_0 e^{-\sigma} \bigg[ g_{ij} \left( - P_1 \mathcal K 
  - \left(2 P'_1 + \half  P_4' + {P'_6}\right) n^k \partial_k \sigma\right)\\
  & -n_i n_j \left( {P'_6   n^k \partial_k \sigma + P_6 \mathcal K } \right)+ P_1 \nabla_{(i} n_{j)} 
  + (2 P'_1+  P'_4 + {2 P'_6}) n_{(i}\partial_{j)} \sigma \bigg] \\
  & + T_0 e^{-\sigma} \bigg[ (P_1+{2P_6}) n_{(i} \partial_{j)} \tilde \delta(f) - (P_1+{P_6}) g_{ij} n^{k}\partial_k \tilde \delta(f)
  -{P_6 n_i n_j n^{k}\partial_k \tilde \delta(f)}\bigg] \Bigg]~~,\\
  {T^{\text{sur}}_{00}} &=  \tilde \delta(f) T_0 e^{\sigma} \bigg[   (- \half P'_4 + {P'_6}) \mathcal K 
  +  P_3 ~n^k\partial_k \sigma
  + \frac{1}{2}  P'_5 ~\epsilon^{ijk} n_i f_{jk} \bigg]
  + \left( - \half T_0 e^{\sigma} P'_4 \right)~ n^k \partial_k \tilde \delta(f)~~,\\
  {T_{\text{sur}}}^i_0 &=  \tilde \delta(f) \bigg[ -T_0^3 e^{-\sigma} P_2 n_j f^{ji} 
  +  T_0 e^{-\sigma}  \epsilon^{ijk} \left( P'_5 n_j \partial_k \sigma \right) \bigg] ~~.
 \end{split}
\end{equation}
This stress tensor must satisfy the conservation equation
\begin{equation} \label{eq:conVV}
\nabla_\mu T^{\mu\nu}_{\text{sur}}=T^{\mu\nu}_{(0)}n_\mu\tilde\delta(f)~~,
\end{equation}
which gives rise to two separate sets of equations as in \S\ref{sec:perfect}, one determining the position of the surface and the other the conservation of the surface stress tensor along surface directions. Indeed, by explicitly using \eqref{STsurPF}, one can verify that the tangential projection of \eqref{eq:conVV} is automatically verified - a trivial consequence of diffeomorphism invariance along the surface.

\subsection{Constraints on surface transport coefficients from the equilibrium partition function}\label{ssec:frstordpartrel} 
%********************************************************************************
%
In this section we write down the next to leading order surface stress tensor in equilibrium by classifying all the terms allowed 
by symmetries. We then reduce this stress tensor along the time circle and compare it with that which follows from the 
partition function. This allows us to see a rich interplay between the surface transport coefficients and the bulk 
second order coefficients. 

\begin{table}
\begin{center}
\begin{tabular}{ |c|c|c| } 
 \hline
 $a$ & $\mathcal S_{(a)}$ & Reduced form $\mathcal S_{(a)}^{\text{red}}$ \rule{0pt}{2.6ex} \\ [0.7ex]
 \hline
 1 & $\tilde \delta(f)~ n^{\mu} \mathfrak{a}_{\mu}$ & $\tilde \delta(f)~n^{i} \partial_i \sigma$ \rule{0pt}{2.6ex}\\   [0.6ex]
 2 & $\tilde \delta(f)~K$ & $\tilde \delta(f) \left( \mathcal K + n^{i} \partial_i \sigma \right)$ \rule{0pt}{2.6ex} \\  [0.6ex]
 3 & $\tilde \delta(f)~n^{\mu} \ell_{\mu}$ & $\tilde \delta(f)~\left( \frac{e^{\sigma}}{2}\right) \epsilon^{ijk} n_i f_{jk}$ \rule{0pt}{2.6ex}\\ [0.6ex]
 4 & $n^\mu \partial_{\mu} \tilde \delta(f)$ & $n^i \partial_{i} \tilde \delta(f)$ \rule{0pt}{2.6ex}\\ [0.6ex]
 \hline
\end{tabular}
\caption{Scalars}\label{Tab:tabscal}
\end{center}
\end{table}

\begin{table}
\begin{center}
\begin{tabular}{ |c|c|c| } 
 \hline
 $a$ & $\mathcal V^{\mu}_{(a)}$ & Reduced form $\mathcal V^{i}_{(a)}$ \rule{0pt}{2.6ex} \\ [0.7ex]
 \hline
 1 & $\tilde \delta(f)~\mathcal P^{\mu \alpha} \mathfrak{a}_{\alpha}$ & $\tilde \delta(f)~\mathcal P^{ij} \partial_j \sigma$ \rule{0pt}{2.6ex}\\   [0.6ex]
 2 & $\tilde \delta(f)~\mathcal P^{\mu \alpha} n^{\nu} \omega_{\nu \alpha}$ 
 & $\tilde \delta(f)~ \left(\frac{e^{\sigma}}{2}\right)\mathcal P^{ij} f_{jk} n^{k}$ \rule{0pt}{2.6ex}\\  [0.6ex]
% % %  3 & $\mathcal P^{\mu \alpha} u^{\beta} \nabla_{\beta} n_{\alpha}$ & $\left(\frac{e^{\sigma}}{2}\right)\mathcal P^{ij} f_{jk} n^{k}$ \\ [0.6ex]
 3 & $\tilde \delta(f)~\mathcal P^{\mu \alpha} n^{\beta} \nabla_{\beta} n_{\alpha}$ 
 & $\tilde \delta(f)~ \mathcal P^{ij} n^k \nabla_k n_j$ \rule{0pt}{2.6ex}\\ [0.6ex] 
 4 & $\tilde \delta(f)~\mathcal P^{\mu \alpha}\ell_{\alpha}$ 
 & $\tilde \delta(f)~ \left(\frac{e^{\sigma}}{2}\right)\mathcal P^{ij} g_{jk} \epsilon^{klm} f_{lm}$ \rule{0pt}{2.6ex}\\ [0.6ex] 
 5 & $\tilde \delta(f)~\mathcal P^{\mu \alpha}\epsilon_{\alpha \lambda  \nu \sigma}  u^{\lambda} n^{\nu} \mathfrak{a}^{\sigma} $ 
 & $\tilde \delta(f)~ \mathcal P^{ij} g_{jk} \epsilon^{klm} n_l \partial_m \sigma$ \rule{0pt}{2.6ex} \\ [0.6ex] 
 6 & $\tilde \delta(f)~\mathcal P^{\mu \alpha}\epsilon_{\alpha \nu \lambda \sigma} u^{\nu} \nabla^{\lambda} n^{\sigma}$ 
 & $\tilde \delta(f)~ \mathcal P^{ij} g_{jk} \epsilon^{klm} \nabla_l n_m$ \rule{0pt}{2.6ex}\\ [0.6ex] 
 7 & $\mathcal P^{\mu \alpha} \partial_{\alpha} \tilde \delta(f)$ 
 & $\mathcal P^{ij} \partial_{j} \tilde \delta(f)$ \rule{0pt}{2.6ex}\\ [0.6ex]
% % % % %  8 & $\mathcal P^{\mu \alpha}\epsilon_{\alpha \nu \lambda \sigma} n^{\nu} u^{\lambda} \nabla^\sigma \tilde \delta(f) $ 
% % % % %  & $\mathcal P^{ij} g_{jk} \epsilon^{klm} n_l \partial_m \tilde \delta(f)$ \rule{0pt}{2.6ex}\\ [0.6ex] 
 \hline
\end{tabular}
\caption{The first order vectors on the surface projected on the surface and orthogonal to the velocities. 
Here $\mathcal P^{\mu \nu} = \mathcal G^{\mu \nu} + u^{\mu} u^{\nu} - n^{\mu} n^{\nu}$, is the projector 
orthogonal to both $u^{\mu}$ and $n^{\mu}$. The spatial components of this projector, projects orthogonal to $n^i$ in 
a given time slice, that is $\mathcal P^{ij} = g^{ij} - n^i n^j$. Also, note that in the reduced language ${\mathcal {V}_{0}}^{(j)} =0$.}\label{Tab:tabvec}
\end{center}
\end{table}

Under the assumption of time independence, at first order on the surface, 
the non-zero linearly independent terms have been classified in Tables \ref{Tab:tabscal}, \ref{Tab:tabvec}
and \ref{Tab:tabten}. The presence of the vectors $u_\mu$ and $n_\mu$ breaks down the local Lorentz symmetry 
at the surface to a smaller subgroup. The classification is based on transformation properties 
of the surface quantities under this preserved subgroup. We refer to the objects as scalars, vectors and 
tensors, depending on their transformation properties under this subgroup.
Note that we have defined $K_{\mu \nu} = \nabla_{(\mu} n_{\nu)}$. 

We would like to point out that
in Table \ref{Tab:tabvec}, we have not included the term $\mathcal P^{\mu \alpha} u^{\beta} \nabla_{\beta} n_{\alpha}$
because upon reduction it evaluates to the same result as $\mathcal P^{\mu \alpha} n^{\nu} \omega_{\nu \alpha}$. Hence, in the 
stationary equilibrium case under consideration, these two terms are not independent. 
Also, owing to the identity 
$$\epsilon^{ijk} n_j \partial_k \tilde \delta(f) = \tilde \delta(f) \epsilon^{ijk}\partial_j n_k~~,$$
the term $\mathcal P^{\mu \alpha}\epsilon_{\alpha \nu \lambda \sigma} n^{\nu} u^{\lambda} \nabla^\sigma \tilde \delta(f) $
is not independent from  $V^{\mu}_{(6)}$. Also note that in Table \ref{Tab:tabten}, $P^{\alpha \beta} K_{\alpha \beta}$ is distinct from
$K$, since $u^{\alpha} u^{\beta} K_{\alpha \beta}$ is non-zero. In the stationary case, 
$P^{\alpha \beta} K_{\alpha \beta}$ reduces to $\mathcal K$, while $K = \mathcal K + n\cdot\partial \sigma$,
as shown in Table \ref{Tab:tabscal}.

\begin{table}
\begin{center}
\begin{tabular}{ |c|c|c| } 
 \hline
 $k$ & $\mathcal T^{\mu \nu}_{(k)}$ & Reduced form $\mathcal T^{ij}_{(k)}$ \rule{0pt}{2.6ex} \\ [0.7ex]
 \hline
 1 & $\mathcal P^{\mu \alpha}  \mathcal P^{\nu \beta}  K_{\alpha \beta} - \frac{1}{2}\mathcal P^{\mu \nu}\mathcal P^{\alpha \beta} K_{\alpha \beta}$ 
 & $\mathcal P^{i k}  \mathcal P^{j m}  \left( \mathcal K_{k m} 
 - \frac{1}{2} g_{km} \mathcal K \right)$ \rule{0pt}{3ex}\\   [1ex]
 \hline
\end{tabular}
\caption{Symmetric traceless tensor. Upon reduction, $\mathcal T^i_0$ and $\mathcal T_{00}$ components 
of the tensor vanishes.}\label{Tab:tabten}
\end{center}
\end{table}

Since we would like to have, the velocity at the surface, to be equal to the bulk velocity evaluated at the surface, there 
is no freedom in choosing a frame at the surface once the bulk frame has been chosen, as discussed in \S\ref{sssec:genpart2}.  
In order to respect the continuity of fluid variables and to naturally impose the condition \eqref{velcond}, we shall proceed 
with the frame choice as described in \S\ref{sssec:genpart2}. This frame choice only constrains the 
form of the bulk stress tensor and leaves the surface stress tensor unconstrained. Therefore, while constructing 
the surface stress tensor at first order, we have to write down all possible terms that are allowed by symmetry 
without imposing any restrictions. We have 
\begin{equation}\label{surPhenoST}
\begin{split}
T_{\text{sur}}^{\mu \nu} =&~ \sum_{i=1}^4 \left( \mathcal P^{\mu \nu} ~{\mathfrak{s1}}_{i} ~\mathcal S_{(i)} +
u^{\mu} u^{\nu} ~{\mathfrak{s2}}_{i} ~\mathcal S_{(i)} + n^{\mu} n^{\nu} ~{\mathfrak{s3}}_{i} ~\mathcal S_{(i)}
+ u^{(\mu} n^{\nu)} ~{\mathfrak{s4}}_{i} ~\mathcal S_{(i)}\right)\\
& + \sum_{i=1}^7 2 \left( \mathfrak{v1}_{i} ~u^{(\mu} \mathcal V^{\nu)}_{(i)}+ \mathfrak{v2}_{i} ~n^{(\mu} \mathcal V^{\nu)}_{(i)}\right) 
+ {\mathfrak {t}} \mathcal T^{\mu \nu}~~,
\end{split}
\end{equation}
where $\{ \mathcal S_{(i)}, \mathcal V^{\nu}_{(i)},  \mathcal T^{\mu \nu} \} $ are specified in the second column of Tables \ref{Tab:tabscal}, \ref{Tab:tabvec}
and \ref{Tab:tabten}, respectively. The corresponding surface transport coefficients are denoted by 
$\{\mathfrak{s1}_i,\mathfrak{s2}_i,\mathfrak{s3}_i,\mathfrak{s4}_i,\mathfrak{v1}_i, \mathfrak{v2}_i,\mathfrak{t}\}$. 
As we may already foresee, among these $4\times4 + 2\times7 + 1 = 31$
transport coefficients, only 3 are independent. The rest are determined in terms one another or bulk second order 
transport coefficients. We will now work out these relations.

If we consider the reduction of \eqref{surPhenoST} along the time direction, we obtain the following reduced stress tensor
\begin{equation}\label{surPhenoSTred}
\begin{split}
&{T^{\text{sur}}}_{00} = \sum_{a=1}^4 e^{2\sigma} ~{\mathfrak{s2}}_{a} ~\mathcal S_{(a)}^{\text{red}} ~~,\\
&{T_{\text{sur}}}_{0}^i =\sum_{a=1}^4 \left( - e^{\sigma} \right) n^i~{\mathfrak{s4}}_{a} ~\mathcal S_{(a)}^{\text{red}} 
+  \sum_{a=1}^7 \left( - e^{\sigma} \right) \mathfrak{v1}_{a} \mathcal V^{i}_{(a)}~~, \\
&{T_{\text{sur}}}^{ij} =\sum_{a=1}^4 \left( \mathcal P^{ij} ~{\mathfrak{s1}}_{a} ~\mathcal S_{(a)}^{\text{red}} 
+ n^{i} n^{j} ~{\mathfrak{s3}}_{a} ~\mathcal S_{(a)}^{\text{red}}\right)
+ \sum_{a=1}^7 \left(  \mathfrak{v2}_{a} ~n^{(i} \mathcal V^{j)}_{(a)}\right) 
+ {\mathfrak {t}} ~\mathcal T^{ij} ~~,\\
\end{split}
\end{equation}
where $\{ \mathcal S_{(a)}^{\text{red}}, V^{i}_{(a)},  \mathcal T^{ij} \} $ are specified in the third column of Tables \ref{Tab:tabscal}, \ref{Tab:tabvec}
and \ref{Tab:tabten}, respectively.

Now comparing \eqref{surPhenoSTred} with \eqref{STsurPF} we get 
\begin{equation} \label{comprel}
\begin{split}
& \mathfrak{s2}_1= T_0 e^{-\sigma} \left( P_3 + \half P_4' - P_6' \right),
~\mathfrak{s2}_2= T_0 e^{-\sigma} \left( - \half P_4' +  P_6' \right) ,
~\mathfrak{s2}_3=T_0 e^{-2\sigma} P_5' ,~\mathfrak{s2}_4= - \frac{T_0 e^{-\sigma}}{2} P_4' , \\
&\mathfrak{s4}_a = 0, ~\forall a \in \{1,2,3,4\}, \mathfrak{v1}_a = 0, ~\forall a \in \{1,4,3,6,7\},
 \mathfrak{v1}_2 = 2 (T_0^3 e^{-3 \sigma}) P_2, 
%  ~  \mathfrak{v1}_6 = \mathfrak{v1}_8 = \half \left(- T_0 e^{-2 \sigma} P_5 \right), 
\mathfrak{v1}_5 =\left(- T_0 e^{-2 \sigma} P_5' \right),  \\
&\mathfrak{v2}_a = 0, ~\forall a \in \{2,4,5,6\}, \mathfrak{v2}_1 = T_0 e^{-\sigma} \left( 2P_1'+P_4'+2 P_6' \right) , 
~\mathfrak{v2}_3 =T_0 e^{-\sigma} P_1, \\ & 
\mathfrak{v2}_7 = T_0 e^{-\sigma} \left( P_1 + 2 P_6 \right),
\mathfrak{s1}_1= T_0 e^{-\sigma} \left( P_1 - \left( 2 P_1' + \half P_4' + P_6' \right) \right), ~\mathfrak{s1}_2= - T_0 e^{-\sigma} P_1,
~\mathfrak{s1}_3 = 0, \\ 
& \mathfrak{s1}_4 = - T_0 e^{-\sigma} \left( P_1 + P_6\right), 
~\mathfrak{s3}_1 = T_0 e^{-\sigma} \left( P_1 + P_6 + \half P_4' \right), 
~\mathfrak{s3}_2 = -T_0 e^{-\sigma} \left( P_1 + P_6 \right), \\ 
& \mathfrak{s3}_3 =  0,~ \mathfrak{s3}_4 = 0, ~\mathfrak{t} = T_0 e^{-\sigma} P_1~~. 
\end{split}
\end{equation}
% % % % & \alpha_1 - \frac{1}{3} \alpha_5 + \gamma_1 + \mathfrak{c} (T_0 e^\sigma P_3) = -\left( 2 P'_1 + \half T_0 e^{-\sigma} P_4 \right),
% % % % ~ -\half \beta_1 e^\sigma + \mathfrak{c} \left( \half e^\sigma P'_5 \right) = 0. \\
% % % % & \gamma_1 - \mathfrak{c} \left( \half T_0 e^{-\sigma} P_4' \right) = -\frac{2}{3} P_1, 
% % % % ~ \varsigma_1 - \mathfrak{c} \left( \half T_0 e^{-\sigma} P_4'\right) = - P_1, \\
% % % % & \alpha_2 + \gamma_2 - \mathfrak{c} \left( T_0 e^\sigma P_3\right) = 0, 
% % % % ~ - \half \beta_2 e^\sigma - \mathfrak{c} \left( \half e^\sigma P'_5\right) = 0, \\ &
% % % %  \gamma_2 + \mathfrak{c} \left( \half T_0 e^{-\sigma} P'_4\right) =0, 
% % % % ~ \varsigma_2 + \mathfrak{c} \left( \half T_0 e^{-\sigma} P'_4\right) = 0 \\ &
% % % % \alpha_3 = 2 P'_1 + T_0 e^{-\sigma} P'_4, ~ \varsigma_3 = P_1, \alpha_4=0, ~ \beta_3 =0, \beta_4 =0, 
% % % % \alpha_5 = P_1, \alpha_6 = 0. 
% % % % \end{split}
% % % % \end{equation}
% % % where we have defined $\mathfrak{c} = \left( \partial_T\chi\right) / \left( T_0^2 \partial_T^2 \chi \right) $. 

Let us recall from \cite{Banerjee:2012iz} that $P_1, P_2$ and $P_3$ may be expressed in terms of the bulk transport coefficients 
in the following way
\footnote{Here we pick up only three specific relations; they can be expressed in several other ways using the 
relations between bulk second order coefficients, as was obtained in \cite{Banerjee:2012iz}.
The bulk second order coefficients in \eqref{prevrel} appeared in the Landau-frame stress tensor in the following way
\begin{equation}\label{STLandau}
\begin{split}
T_{\mu \nu} =& T \left( \ \kappa_1 ~\tilde R_{\langle\mu \nu\rangle} + \kappa_2 ~K_{\langle\mu \nu\rangle} + \lambda_3  \ 
\omega_{\langle\mu}^{\,\, \alpha} \omega_{\alpha \nu\rangle} + \lambda_4 \ \mathfrak{a}_{\langle \mu}\mathfrak{a}_{\nu \rangle} 
 + P_{\mu \nu}(\zeta_2  \tilde R+\zeta_3  \tilde R_{\alpha \beta}u^\alpha u^\beta +\xi_3 \omega^2+\xi_4 \mathfrak{a}^2 ) \ \right)~~.
\end{split}
\end{equation}
For further details of the conventions, we refer the reader to \cite{Banerjee:2012iz}. In the 
orthogonal-Landau-frame as defined in \S \ref{sssec:genpart2}, which is the most suitable bulk frame for 
describing the fluid configurations with a surface, the stress tensor takes the form \eqref{STOrthLandau}. Note that the bulk 
transport coefficients appearing in \eqref{prevrel} have very similar physical meanings in both the frames. See appendix \ref{app:Frametrans} for further 
details.}
\begin{equation}\label{prevrel}
 P_1 = \kappa_1, ~ P_2 = \frac{1}{8 T^2} ~\left(  2 \kappa_1 + \kappa_2 - \lambda_3 \right), 
 ~ P_3 = \frac{T P_{TT}}{P_T}  \left( \frac{2}{3} \left(\kappa_2 - \kappa_1 \right) - 2 \zeta_2 + \zeta_3\right)~~.
\end{equation}
Finally, eliminating the $P_i$ variables from \eqref{comprel}, we can summarize the following relations involving surface first order 
coefficients and second order bulk transport coefficients
\begin{equation} \label{mainconst}
\begin{split}
&\mathfrak{s4}_a = 0, ~\forall a \in \{1,2,3,4\},~~ \mathfrak{v1}_a = 0, ~\forall a \in \{1,3,4,6,7\}~~, \\
&\mathfrak{v2}_a = 0, ~\forall a \in \{2,4,5,6\},~~ \mathfrak{s1}_3 = 0,~~\mathfrak{s3}_3 =  0,~ \mathfrak{s3}_4 = 0~~,\\
&\mathfrak{s3}_1 = -(\mathfrak{s1}_4+\mathfrak{s2}_4),~~ \mathfrak{s3}_2 = \mathfrak{s1}_4,~~ \mathfrak{s1}_2 = - \mathfrak{t}, 
~~  \mathfrak{s2}_3 = - \mathfrak{v1}_5~~, \\
& \mathfrak{s2}_2 =  \mathfrak{s2}_4 + T \partial_T  \left( \mathfrak{s1}_4 + \mathfrak{t} \right) - \left( \mathfrak{s1}_4 + \mathfrak{t}\right), 
~~ \mathfrak{v2}_1 = 2 \left( T \partial_T \mathfrak{s1}_4 - \mathfrak{s1}_4 - \mathfrak{s2}_4   \right)~~, \\
& \mathfrak{v2}_3 = -(2\mathfrak{s1}_4 + \mathfrak{v2}_7),~~ \mathfrak{v2}_7 = -\mathfrak{t}-2 \mathfrak{s1}_4,~~
 \mathfrak{s1}_1 + \mathfrak{s3}_1 + T \partial_T \left( \mathfrak{s1}_4 - \mathfrak{t} \right) = 0  ~~, \\
 & \mathfrak{t} = T \kappa_1,~~ \mathfrak{v1}_2={T \over 4}(2 \kappa_1+\kappa_2-\lambda_3), 
 ~~ \mathfrak{s2}_1 = {P_{TT}T^2 \over P_T} \left({2\over 3} (\kappa_2 - \kappa_1)-2 \zeta_2+\zeta_3\right)-\mathfrak{s2}_2~~.
\end{split}
\end{equation} 
These relations \eqref{mainconst} are one of the central results of the paper. Let us now highlight some of the most interesting 
aspects of these relations \eqref{mainconst}. The last three relations in \eqref{mainconst} relate bulk transport coefficients to 
those in the boundary. This shows that the linear response to particular deformation of the surface is intimately related to 
some, otherwise unrelated, transport coefficient in the bulk. Particularly interesting is the fact that  $\mathfrak{t}$ and 
$\kappa_1$ are proportional to each other. This physically implies that the linear response to a longitudinal stretch of the 
surface is entirely determined by how the fluid reacts to a change in background curvature. 

Another noteworthy fact is that the parity odd term introduced in \eqref{surpartfrstord}, is reminiscent 
of the possible parity odd term in $2+1$ dimensional space-filling fluids discussed in \cite{Banerjee:2012iz, Jensen:2012jh}.
It leads to two non-zero parity odd coefficients, namely $\mathfrak{s2}_3$ and $\mathfrak{v1}_5$. 
The scalar $\mathcal S_2$ is proportional to $n\cdot \ell$, while $\mathcal V^\mu_5$ is non-zero only when 
the acceleration at the surface $\mathfrak{a}_\mu$ has a component parallel to the surface (see Tables \ref{Tab:tabscal} and \ref{Tab:tabvec}).
It is interesting to note that although, space-filling uncharged fluids do not have any parity odd stationary transport at 
next to leading order, such a transport may exist at the surface of a finite lump of the same fluid.

Some of these constraints can be anticipated from the structural aspects of the 
conservation equation \eqref{eq:conVV} on an arbitrary surface stress tensor, as explained in Appendix \ref{ssec:aaa}. 
In Appendix \ref{sec:entropy}, the remaining constraints are also obtained through an entropy current argument, 
particularly adapted to deal with the stationary transport coefficients.

%***************************************************************************************************************************
\subsection{Description in terms of original fluid variables}\label{ssec:frstordAc}
%***************************************************************************************************************************
In this section we lift the partition function of stationary neutral fluids \eqref{surpartsecord} to a four-dimensional covariant action.\footnote{Note that we are using the terminology \emph{action} in a slightly different way than \cite{Bhattacharya:2012zx}. This is because in the presence of a surface we can view \eqref{eq:action} as an action functional for the surface $f$. Indeed, the surface part of \eqref{eq:action}, to leading order, is equivalent to the DBI action for co-dimension one branes when $\chi$ is constant and no worldvolume or background fields are present.} This action assumes that existence of a spacetime Killing vector field $\textbf{k}^{\mu}$ with modulus $\textbf{k}=\sqrt{-\mathcal{G}_{\mu\nu}\textbf{k}^{\mu}\textbf{k}^{\nu}}$ along which the fluid flows are aligned, i.e., $u^{\mu}=\textbf{k}^{\mu}/\textbf{k}$ as in \cite{Jensen:2012jh, Armas:2013hsa}. 
% \footnote{This lifting is not possible to do in a straightforward way for charged fluids in the presence of anomalies \cite{Banerjee:2012iz}.} 
We also assume that the surface is characterized by the same bulk Killing vector field restricted to the surface such that $\textbf{k}|_{f=0}=\sqrt{-\mathcal{H}_{ab}\textbf{k}^{a}\textbf{k}^{b}}$, where $\mathcal{H}_{ab}$ is the induced metric on the surface. Generically, we may write the effective action as the sum of a bulk and surface parts,
\begin{equation} 
\mathcal{I}=\int_{\mathcal{M}}\sqrt{-\mathcal{G}}~\mathcal{I}_{\text{blk}}\left(\mathcal{G}_{\mu\nu},\partial\mathcal{G}_{\mu\nu},...\right)+\int_{\partial\mathcal{M}}\sqrt{-\mathcal{H}} ~\mathcal{I}_{\text{surf}}\left(\mathcal{H}_{ab},\partial \mathcal{H}_{ab}, K_{ab},...\right)~~.
\end{equation}
This effective action, for neutral fluids up to second order, as in the case of the partition function, is described in terms of six transport coefficients,
\begin{equation} \label{eq:action}
\mathcal{I}=\int_{\mathcal{M}}\sqrt{-\mathcal{G}} \left(P+ \tilde P_1 \tilde R+\tilde P_2 \omega^2+\tilde P_3 \mathfrak{a}^2\right)+\int_{\partial\mathcal{M}}\sqrt{-\mathcal{H}} \left(\chi+\tilde{\mathcal{B}}_1 \mathfrak{a}^{\mu}n_\mu+\tilde{\mathcal{B}}_2\ell^{\mu}n_\mu+\tilde{\mathcal{B}}_3 K\right)~~,
\end{equation} 
where $\tilde R$ is the Ricci scalar of the spacetime, $\omega^2=\omega_{\mu\nu}\omega^{\mu\nu}$ and $\mathfrak{a}^2=\mathfrak{a}^{\mu}\mathfrak{a}_\mu$. Here, the pressure, the surface tension and all transport coefficients are functions of the local fluid temperature $T$ which is given in terms of the global temperature $T_0$ and the modulus $\textbf{k}$ via the relation $T_0=\textbf{k}T$.

The bulk part of this action has been written down in \cite{Jensen:2012jh}, and the coefficients $\tilde P_i$ measure the response of the fluid to background curvature, vorticity and acceleration. The surface part of this action was analyzed in \cite{Armas:2013hsa} in arbitrary spacetime dimensions. However, there, as explained in Appendix~\ref{ssec:aaa}, since the bulk pressure vanished at the surface, the scalars $ \mathfrak{a}^{\mu}n_\mu$ and $K$ were not independent. Here the response coefficient $\tilde{\mathcal{B}}_3$ is the surface modulus of rigidity of the surface fluid \cite{Armas:2013hsa} while $\tilde{\mathcal{B}}_1$ encodes the response to centrifugal acceleration on the surface. Furthermore, dimension-dependent scalars were not analyzed in \cite{Armas:2013hsa}. In this case, the scalar $\ell^{\mu}n_\mu$ is well known from the study of $2+1$ parity odd fluids \cite{Jensen:2012jh} and encodes the response due to vorticity at the surface. Despite being written in four spacetime dimensions, the action \eqref{eq:action} generalises to arbitrary spacetime dimension with $\tilde{\mathcal{B}}_2=0$. 

The equations of motion can be derived from the action \eqref{eq:action} by performing a general diffeomorphism of the form $\delta \mathcal{G}_{\mu\nu}=2\nabla_{(\mu}\xi_{\nu)}$ and decomposing $\xi_\mu$ into tangential and normal components to the surface such that $\xi_\mu={e_{\mu}}^{a}\xi_{a}+n_\mu n^\nu\xi_{\nu}$ as in \cite{Armas:2013hsa}. The surface part of the variation of the action \eqref{eq:action} yields
\begin{equation}
\delta_\xi\mathcal{I}=\int_{\partial\mathcal{M}}\sqrt{-\mathcal{H}}\left(-T^{\mu\nu}_{(0)}n_{\mu}\xi_\nu+\mathbb{T}^{ab}_{\text{sur}}\delta_{\xi} \mathcal{H}_{ab}+n_\rho T^{ab\rho}_{(2)}\delta_{\xi} K_{ab}\right)~~,
\end{equation}
where $\delta_{\xi}$ denote variations along the co-vector field $\xi_{\mu}$ and where we have defined
\begin{equation}
\mathbb{T}^{ab}_{\text{sur}}=\frac{2}{\sqrt{-\mathcal{H}}}\frac{\delta \mathcal{I}}{\delta_\xi \mathcal{H}_{ab}}~~,~~n_\rho T^{ab\rho}_{(2)}=\frac{2}{\sqrt{-\mathcal{H}}}\frac{\delta \mathcal{I}}{\delta_\xi K_{ab}}~~.
\end{equation}
This variation leads to two sets of equations of motion \cite{Armas:2013hsa}, which can equivalently be obtained from \eqref{eq:conVV}. One expresses conservation of the surface stress tensor in directions tangential to the surface, 
\begin{equation} \label{eq:conservation}
 \nabla_{a}\mathbb{T}^{ab}_{\text{sur}}=n_{\rho}T^{ac\rho}_{(2)}\nabla^{b}K_{ac}-2\nabla_a\left(n_\rho T^{ac\rho}_{(2)}{K_{c}}^{b}\right)+T^{\mu\nu}_{(0)}n_\mu e_{\nu}^{b}~~,
\end{equation}
and is automatically satisfied for the stress tensor obtained for each contribution in \eqref{eq:action}. Indeed, the stress tensor \eqref{surPhenoST} with the coefficients \eqref{comprel} satisfies \eqref{eq:conservation}. The other equation is the modified Young-Laplace equation, due to the presence of $\partial_\rho \tilde\delta(f)$ corrections in the surface stress tensor,
 \begin{equation}\label{eq:mYL}
\mathbb{T}_{\text{sur}}^{ab}K_{ab}=n_\rho\nabla_a\nabla_bT^{ac\rho}_{(2)}+n_\rho T^{ab\rho}_{(2)}n_\lambda n^{\mu}{{\tilde{R}}^{\lambda}}_{~a\mu b}+T^{\mu\nu}_{(0)}n_\mu n_\nu~~,
\end{equation}
where the stress tensor $\mathbb{T}_{\text{sur}}^{ab}$, obtained by directly varying \eqref{eq:action} with respect to the induced metric on the surface, is given in terms of the components \eqref{STschC} as\footnote{This expression was derived in \cite{Armas:2013hsa} but using other conventions for the stress tensor. Here we have written it using the conventions in \eqref{STschC} which required using a result from \cite{Vasilic:2007wp}.}
\begin{equation} \label{eq:Tabc}
\mathbb{T}_{\text{sur}}^{ab}=T^{ab}_{(1)}-n^{\lambda}\nabla_\lambda\left(n_\rho T^{\mu\nu\rho}_{(2)}\right){e^{a}}_\mu{e^{a}}_\nu+n_\rho T^{ab\rho}_{(2)} K+T^{abc}_{(2)}v_c-2n_\rho T^{(ac\rho}_{(2)}{K^{b)}}_c~~,
\end{equation}
where $\tilde R_{\mu\nu\lambda\rho}$ is the Riemann tensor of the spacetime and where we have defined $v_{c}={u_c}^{\rho}n^{\lambda}\nabla_\lambda n_\rho$. Note that if we do not consider higher order corrections, then $T^{\mu\nu\rho}_{(2)}$ vanishes and $T^{ab}_{(1)}$ takes the perfect fluid form. In this case, equations \eqref{eq:conservation} and \eqref{eq:mYL} reduce to \eqref{bdystcon} and \eqref{tangential}.

In order to understand the relation between these transport coefficients and the ones appearing in the partition function of Sec.~\ref{ssec:frstordpartrel}, we reduce \eqref{eq:action} over the time circle and obtain
\begin{equation}
\begin{split}
\mathcal W = \log \mathcal Z =&  \int_{\mathcal{M}_s} d^3x \sqrt{g} \frac{e^\sigma}{T_0}\left( P + \tilde P_1 R +\frac{e^{2\sigma}}{4}\left(\tilde P_1+\tilde P_2\right) f_{ij}f^{ij} 
 + \left(\tilde P_3+2\tilde P_1'\right)( \partial \sigma)^2 \right)
 \bigg] \\
&+ \int_{\partial {\mathcal{M}_s}} ~d^2x \sqrt{\gamma} \frac{e^\sigma}{T_0}  \left(  \chi+ \left(\tilde{\mathcal{B}}_1+\tilde{\mathcal{B}}_3-2\tilde P_1\right) n^i\partial_i \sigma 
 +\frac{e^{\sigma}}{2} \tilde{\mathcal{B}}_2 \epsilon^{ijk} n_i f_{jk} + \tilde{\mathcal{B}}_{3}\mathcal{K}\right) \bigg|_{f=0}.
 \end{split}
\end{equation}
 Comparison with \eqref{surpartsecord} leads to the identification of the pressure and surface tension as $\mathcal{P}=P$ and $\mathcal{C}=\chi$ as well as to the relations between higher and lower dimensional transport coefficients. More precisely, we find
\begin{equation}
\begin{split}
&P_1T=-2\tilde P_1~~,~~P_2 T^3=-\frac{\tilde P_1+\tilde P_2}{2}~~,~~P_3 T=-2(\tilde P_3+2\tilde P_1')~~, \\
&\mathcal{B}_1=\tilde{\mathcal{B}}_1+\tilde{\mathcal{B}}_3-2\tilde P_1~~,~~{\mathcal{B}}_2=\frac{e^{\sigma}}{2}\tilde{\mathcal{B}}_2~~,~~{\mathcal{B}}_3=\tilde{\mathcal{B}}_3~~,
\end{split}
\end{equation}
and, if written in the form \eqref{surpartsecord2}, then we can readily identify
\begin{equation}
T P_4'=-2\left(\tilde{\mathcal{B}}_1+\tilde{\mathcal{B}}_3-2\tilde P_1\right)~~,~~TP_5=-e^{\sigma}\tilde{\mathcal{B}}_2~~,~~TP_6=\tilde{\mathcal{B}}_3~~.
\end{equation}
By using the identifications above in \eqref{mainconst} and computing \eqref{eq:Tabc}, one can explicitly check that equation \eqref{eq:conservation} is automatically satisfied.

%**********************************************************************************************************
\section{Fluid configurations in 2+1 dimensions}\label{sec:config}
%**********************************************************************************************************

In this section, we shall construct a few simple stationary fluid configurations to demonstrate the relevance of 
a non-zero surface entropy. The modifications of the fluid equations, in particular the Young-Laplace 
equation \eqref{genLY}, when there is a non-zero surface entropy or equivalently 
a non-trivial dependence of surface tension on temperature, was discussed in \S \ref{sec:perfect}. 
Here, we will work out some particular solutions to these equations and explore 
the consequences of a non-zero surface entropy on the phase diagram of these fluid configurations. 

We will keep our focus only on perfect fluids in $2+1$ dimensions, ignoring possible higher derivative corrections. 
For working out explicit configurations we need the knowledge of the equation of state, for which 
we have to consider a specific system. For this purpose, we will consider the system of 
localized deconfined plasma of $\mathcal N=4$ Yang-Mills theory, compactified down to $2+1$ on a Scherk-Schwarz circle, 
dual to rotating black holes and black rings in Scherk-Schwarz compactified 
$AdS_5$, via the AdS/CFT correspondence \cite{Aharony:2005bm, Lahiri:2007ae}. 

Here, we shall revisit the analysis of \cite{Lahiri:2007ae} in order to find out how the results there are 
modified in the presence of non-zero surface entropy. In this section, we would like to 
briefly present our main results and therefore refer the reader to \cite{Lahiri:2007ae}
and the references therein (also see \cite{Bhardwaj:2008if, Bhattacharya:2009gm}), for the background material. 

One of the key assumptions in \cite{Lahiri:2007ae} was that the surface tension was constant, which 
we would like to relax here. This was assumed mainly 
because the value of surface tension for the interface between the confined 
and deconfined phase of $\mathcal N =4$ Yang-Mills was only known
at the critical temperature \cite{Aharony:2005bm}. The full dependence of surface tension
on temperature for this interface is still unknown, but we would like to 
parameterize this ignorance in a suitable way and study its consequences. 

%****************************************************************************************************
\subsection{Equation of state and thermodynamic quantities} \label{ssec:configeos}
%****************************************************************************************************
%
The configurations that we shall deal with here (these are the ones 
that were already found in \cite{Lahiri:2007ae} in $2+1$ dimensions), have the feature that 
the temperature is constant throughout the surface of the configuration. 
Also there is an empirical fact that the surface tension 
must decrease with the increase in temperature. This expectation follows from the fact 
that otherwise the surface entropy would become negative. Taking these two 
observations into consideration, and assuming that the surface temperature is very near to 
(slightly above) the phase transition temperature $T_c$, we can consider the following 
dependence of the surface tension on temperature 
\begin{equation}\label{surtempform} 
 \chi(T) = \chi_0 \left(1 + \kappa \left( \frac{T_c - T}{T_c} \right) + \mathcal O  \left( \frac{T_c - T}{T_c} \right)^2  \right)~~,
\end{equation}
where 
%  $T_{s}$ is the temperature at the surface of the configuration and 
$\chi_0$ is 
the value of surface tension at $T_{c}$ 
\footnote{In \cite{Aharony:2005bm}, it was found that at the critical temperature $\chi_0 = 2.0 \left(\frac{\pi^2 N^2}{2}\right) T_c^2$ 
for the plasma-balls in $\mathcal{N} = 4$ SYM.}.
We would like to emphasize that for the configurations of interest, the value 
of temperature at the fluid surface is in general different from the critical temperature. But, as we shall demonstrate 
later in this section, for all the configurations that we consider here, the surface temperature remains very close to $T_c$,
thus justifying our assumption. Also for the 
metastable plasma configurations to exist we must have that $T_s \geq T_c$. As it will turn out, this condition will 
play a very important role in our analysis. 

In \eqref{surtempform}, $\kappa$ is the
parameter that parameterizes our ignorance about the exact temperature dependence 
of surface tension. Positivity of surface entropy implies that $\kappa$ must be positive. 
The results in \cite{Lahiri:2007ae} were obtained with $\kappa = 0$ and the main goal in this section is to 
study how those results are modified as we turn on $\kappa$.

Using the thermodynamical relations from \S \ref{sec:perfect}, we find that the surface energy density $\chi_E$ and the surface entropy $\chi_{S}$
are respectively given by 
\begin{equation}\label{surengent}
 \chi_E \big|_{T_s} = (1 + \kappa) \chi_0~~ , ~~~\chi_{S} \big|_{T_s} = - \frac{\partial \chi}{\partial T} = \kappa \frac{\chi_0}{T_c}~~.
\end{equation}

In addition, we also need the explicit form of the equation of state of the bulk fluid. Following \cite{Lahiri:2007ae}, we take this to be
\begin{equation}\label{eosblk}
 P = \alpha ~T^4 - \rho_0 = \frac{\rho}{3} - \frac{4 \rho_0}{3}~~,
\end{equation}
where $\rho$ is the energy density and $\rho_0$ is the shift in the free energy due to the 
Scherk-Schwarz compactification (see \cite{Lahiri:2007ae}, for more details). 
The other thermodynamic quantities simply follow from \eqref{eosblk} via 
the thermodynamic relations. In particular, the bulk entropy density and temperature 
can be expressed in terms of the energy density as follows 
\begin{equation}\label{blkenttemp}
 s = 4 \left(\frac{\alpha}{3^3}\right)^{\frac{1}{4}} \left( \rho - \rho_0 \right)^\frac{3}{4}~~,
 ~~ T = \left( \frac{\rho -\rho_0}{3 \alpha} \right)^{\frac{1}{4}}~~.
\end{equation}
%
%
%
%****************************************************************************************************
\subsection{Spinning ball and ring} \label{ssec:configeos}
%****************************************************************************************************
%
% Now we will report the $\kappa$ dependence of the rigidly spinning configurations in \cite{Lahiri:2007ae}. 
Before proceeding and state our results, we will briefly mention our conventions, while 
the rest of the details can be checked in \cite{Lahiri:2007ae}. 
The fluid configurations are in $2+1$ dimensional flat space with metric 
\begin{equation} \label{eq:ds}
 ds^2 = -dt^2 + dr^2 + r^2 d\phi^2~~.
\end{equation}
We seek rigidly rotating fluid configurations with the velocity vector $u^{\mu} = \gamma \{1,0,\omega\}$ and 
$\gamma = \left( 1- r^2 \omega^2\right)^{1/2}$. As an ansatz for rigidly rotating stationary configurations, the 
surfaces are are taken to be constant $r$ slices in the spacetime \eqref{eq:ds}.
\footnote{The function $f$ in the previous sections, defining the surface, is taken to be $f = r_+ - r$ for the outer 
surface and $f=r-r_-$ for the inner surface.}

Since the bulk fluid equations are not affected by $\kappa$, the solution in the bulk remains identical to \cite{Lahiri:2007ae}. 
The energy density in the bulk of the fluid has the form 
\begin{equation}\label{rhosolbkl}
 \rho(r) = \rho_0 + \frac{C}{\left( 1- r^2 \omega^2\right)^2}~~,
\end{equation}
$C$ being a constant of integration.
At the inner and outer surfaces (denoted by the subscript $-$ and $+$ respectively), the Young-Laplace equation enforces 
\begin{equation}\label{configbdycond}
 P_\pm =  \pm \chi_0 \left( \frac{1}{r_{\pm}} - \kappa \frac{r_\pm~ \omega^2}{1 - r_\pm^2 ~\omega^2}\right)~~.
\end{equation}
For the rotating balls, there is no inner surface and therefore no condition associated with it. The second term in 
\eqref{configbdycond} is precisely the acceleration term in the modified Young-Laplace equation \eqref{genLY}. 
% \JB{Check the sign here once more}. 
This additional term, in this boundary condition, is one of the two new modifications in our analysis compared to that in \cite{Lahiri:2007ae}.

Now we proceed and obtain the phase diagram for the rotating balls and rings. We wish to plot the total entropy versus 
the total angular momentum at fixed total energy, for these configurations. The total energy $E$ and the total 
angular momentum $L$ is simply obtained by integrating the $T^{tt}$ and $r^2 T^{t\phi}$ components of the stress tensor
\eqref{fullstress}. 

The total entropy $S$, is obtained by integrating the time component of the entropy current 
$J^{\mu}_{s} = \mathfrak{s} ~u^{\mu}$. This is equivalent to integrating $\gamma \mathfrak{s}$, where $\mathfrak{s}$
is the total entropy density, including contributions from the surface
\begin{equation}\label{configentden}
  \mathfrak{s} = s\theta(f) + \sum_i \chi_s ~\delta(f_i)~~,
\end{equation}
where $s$ is given by \eqref{blkenttemp} while $\chi_s$ is given by \eqref{surengent}. 
This inclusion of the surface contribution to \eqref{configentden} is 
the second important modification in our analysis. 

Following \cite{Lahiri:2007ae}, we introduce dimensionless quantities 
\begin{equation}
 \tilde E  = \left( \frac{\rho_0}{\pi \chi_0^2}\right) E ~~, ~~ \tilde L = \left(\frac{\rho_0^2}{\pi \chi_0^3} \right)L~~,
 ~~ \tilde S = \left(\frac{\rho_0^{5/4}}{\pi \alpha^{1/4} \chi_0^2}\right)S~~,
 ~~\tilde \omega = \left( \frac{\chi_0}{\rho_0}\right)\omega~~,~~ \tilde r = \left( \frac{\rho_0}{\chi_0}\right)\omega~~,
\end{equation}
where we have also defined a velocity $v = \tilde r \tilde \omega$.
For the rotating ball we have 
\begin{equation}\label{ballels}
 \begin{split}
  \tilde E &=  \frac{1}{\tilde \omega^2} \left( 4 v_+^2 - v_+^4 + (5+2\kappa) \ \tilde \omega v_+ - (1+\kappa)\ \tilde \omega v_+^3 \right)~~,\\
  \tilde L &=  \frac{2}{\tilde \omega^3} \left(v_+^4 +	(1+\kappa)\tilde \omega v_+^3\right)~~,\\
  \tilde S &=  \frac{4 v_+^{5/4}}{\left( 1-v_+^2\right)^{1/4} \tilde \omega^2}\left (-v_ + ^3 - (1+\kappa)  v_ + ^2 \tilde \omega 
   + v_ + +\tilde \omega^2 \right)^{3/4} + \frac{2 \kappa v_+}{ \tilde \omega  \left( 1-v_+^2\right)^{1/2}}~~,
% %   \frac{2 v_+^2}{\tilde \omega^2} \sqrt{1-v_+^2} \left(1 + \frac{\tilde \omega}{v_+} \right)^{\frac{3}{4}}\\
 \end{split}
\end{equation}
where $v_+$ is the velocity at the surface of the ball. 
In turn, for the rotating ring we have 
\begin{equation}\label{ringels}
 \begin{split}
  \tilde E &=  \frac{1}{\tilde \omega^2} \left( 4 (v_+^2-v_-^2) - (v_+^4-v_-^4) + (5+2\kappa) \ \tilde \omega (v_+ + v_-) 
  - (1+\kappa)\tilde \omega (v_+^3+v_-^3) \right)~~,\\
  \tilde L &= \frac{2}{\tilde \omega^3} \left((v_+^4-v_-^4) + (1+\kappa)\tilde \omega (v_+^3+v_-^3)\right)~~,\\
  \tilde S &=  \frac{4 v_+^{5/4}}{\left( 1-v_+^2\right)^{1/4} \tilde \omega^2}\left (-v_ + ^3 - (1+\kappa)  v_ + ^2 \tilde \omega 
  + v_+ +\tilde \omega^2 \right)^{3/4} + \frac{2 \kappa v_+}{ \tilde \omega  \left( 1-v_+^2\right)^{1/2}} ~~,\\
 & -\frac{4 v_-^{5/4}}{\left(1-v_-^2 \right)^{1/4} \tilde \omega^2}\left (-v_- ^3 - (1+\kappa)  v_- ^2 \tilde \omega 
  + v_- +\tilde \omega^2 \right)^{3/4} 
   + \frac{2 \kappa v_-}{ \tilde \omega  \left( 1-v_-^2\right)^{1/2}}~~,
% %   \frac{2}{\tilde \omega^2} \left(\sqrt{1-v_+^2} \left(1 + \frac{\tilde \omega}{v_+} \right)^{\frac{3}{4}} 
% %   - \sqrt{1-v_-^2} \left(1 - \frac{\tilde \omega}{v_-} \right)^{\frac{3}{4}}\right) \\
 \end{split}
\end{equation}
where $v_+$ and $v_-$ are respectively the velocities at the outer and inner surface of the ring. 
We must point out that $v_+$, $v_-$ and $\tilde \omega$ are not independent parameters for the rings. In fact, they must be
related by the condition that the following two functions, must be identical
\begin{equation}\label{gfuncs}
\begin{split}
 g_+ &= \left(1-v_+^2\right) \left( \left(1-v_+^2\right)  + \left(1-(1+\kappa)v_+^2\right) \frac{\tilde \omega}{v_+} \right)~~, \\
 ~ g_- &= \left(1-v_-^2\right) \left( \left(1-v_-^2\right)  - \left(1-(1+\kappa)v_-^2\right) \frac{\tilde \omega}{v_-} \right)~~.
 \end{split}
\end{equation}
As expected,
these expressions reduce to their counterparts in \cite{Lahiri:2007ae} when we set 
$\kappa = 0$. 

%****************************************************************************************************
\subsection{Phase diagram for spinning balls and rings} \label{ssec:configeos}
\begin{figure}[!tbp]
  \centering
  \begin{minipage}[b]{0.4\textwidth}
    \includegraphics[width=\textwidth]{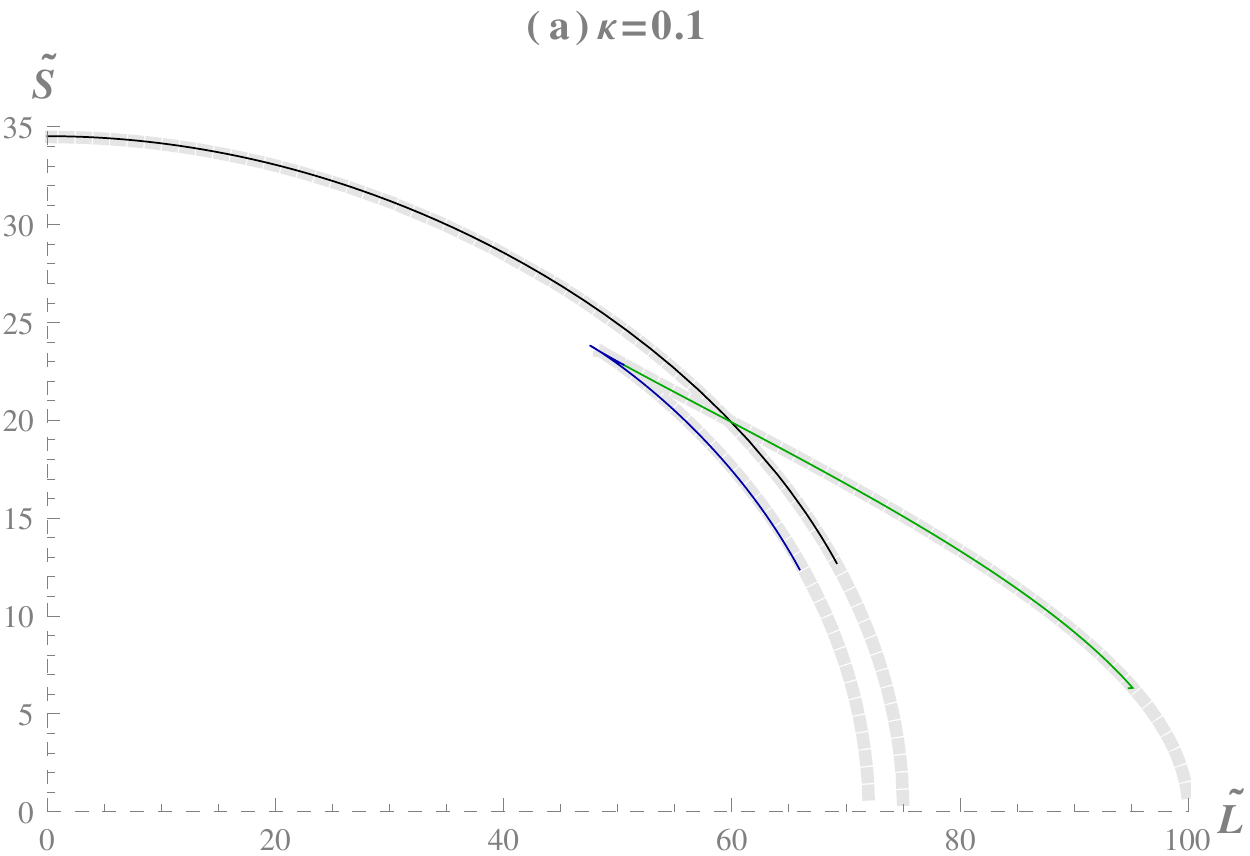}
%     \caption{Flower one.}
  \end{minipage}
  \hfill
  \begin{minipage}[b]{0.4\textwidth}
    \includegraphics[width=\textwidth]{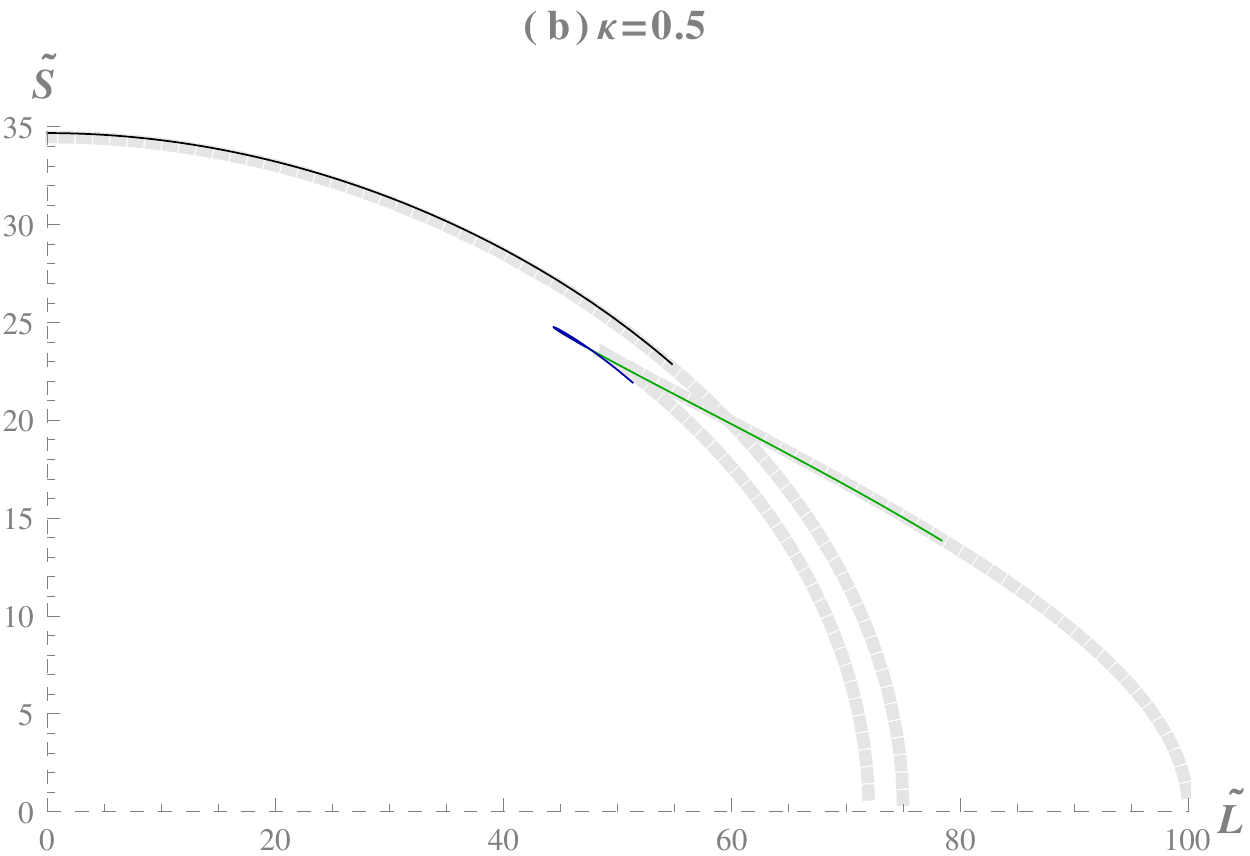}
%     \caption{Flower two.}
  \end{minipage}
  \caption{Phase diagram for the rotating ball and ring configurations for $\tilde E = 40$. The blue line refers to the fat 
  ring while the green line refers to the thin ring.}\label{fig:pdgm}
\end{figure}

%****************************************************************************************************
% 
The phase diagram that emerges out of \eqref{ballels} and \eqref{ringels} has been plotted in Fig. \ref{fig:pdgm}. We have plotted the 
total entropy $\tilde S$ versus total angular momentum $\tilde L$ for a fixed total energy $\tilde E$. 
We have used the same fixed value of energy $\tilde E = 40$, as in \cite{Lahiri:2007ae}, so as to facilitate easy comparison. In
fact, in both the plots in Fig. \ref{fig:pdgm}, we have displayed the $\kappa =0$ phase diagram with light gray lines. 

We have displayed the phase diagram for two values of $\kappa = 0.1, 0.5$. The dark line represents the rotating plasma-ball solution 
while the blue and the green lines represent the rotating fat and thin plasma-ring respectively. 

The main qualitative difference that we find here, as compared to \cite{Lahiri:2007ae}, is that neither the spinning ball nor the rings 
reach zero entropy. This is because, at a given energy $\tilde E$, there is an upper bound for the velocity at the outer surface $v_+$, 
which lies below $1$, for non-zero $\kappa$. 
This upper bound on velocity is the point, where the curves terminate, while the curves for $\kappa =0$ continue 
to zero entropy as $v_+$ approaches $1$. 

\begin{figure}[!tbp]
  \centering
  \begin{minipage}[b]{0.4\textwidth}
    \includegraphics[width=\textwidth]{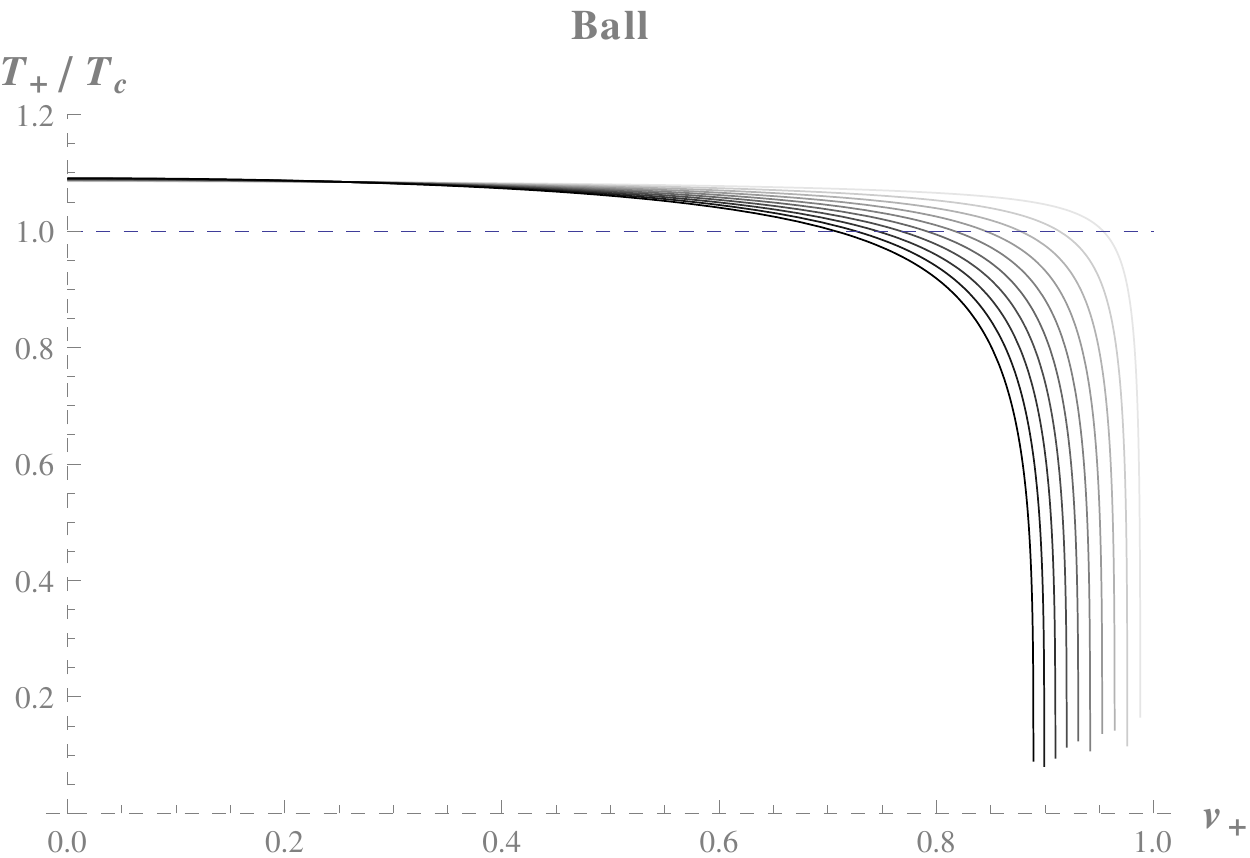}
%     \caption{Flower one.}
  \end{minipage}
  \hfill
  \begin{minipage}[b]{0.4\textwidth}
    \includegraphics[width=\textwidth]{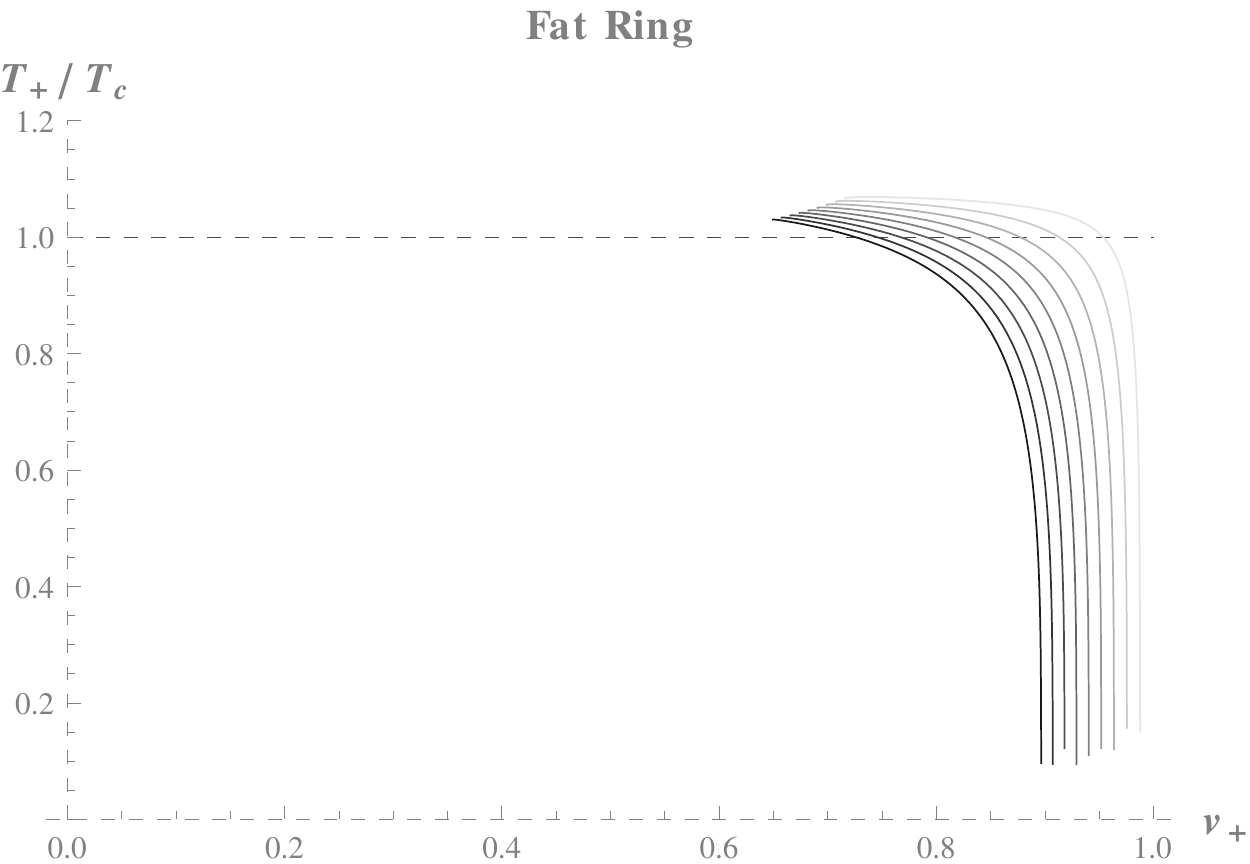}
%     \caption{Flower two.}
  \end{minipage}
  \caption{Temperature at the outer surface of the rotating ball and the fat ring as a function of the velocity 
  at the outer surface $v_+$, at fixed energy $\tilde E = 40$. 
%   For the thin ring, the behaviour of temperature is 
%   identical to that of the fat ring. 
  The various lines represent values of $\kappa$, ranging from $0.1$ to $0.9$, the darkest 
  line corresponding to $0.9$.}\label{fig:Tempplots}
\end{figure}

This bound on $v_+$ arises from the fact that the temperature at the outer surface $T_+$,
reaches the phase transition temperature $T_c$ at the upper bound for $v_+$. At higher values of $v_+$, even if it remains below $1$, 
the temperature at the surface would drop below $T_c$ and the configuration would cease to exist. We have demonstrated this behaviour 
of the surface temperature $T_+$, in Fig. \ref{fig:Tempplots}.

In Fig.\ref{fig:Tempplots}, the temperature at the outer surface of the rotating ball and the fat ring have been plotted as a function of the velocity 
at the outer surface $v_+$ at fixed energy $\tilde E = 40$. For the thin ring, the behaviour of temperature is 
identical to that of the fat ring. The various lines represent values of $\kappa$ ranging from $0.1$ to $0.9$, where the darkest 
line corresponds to $0.9$. As it is apparent from Fig.\ref{fig:Tempplots}, the value of $v_+$ for which the temperature dips 
below the dotted blue line, representing the phase transition temperature, decreases with the increase in $\kappa$. 

For the rings, there is also a lower bound on $v_+$, below which the solutions ceases to exist. This was also present for $\kappa =0$.
Also, as it is apparent from Fig. \ref{fig:Tempplots}, the surface temperature for all the configurations 
remains very close to $T_c$. This justifies 
our initial assumption that, in this analysis, we have taken the value of the surface tension and surface entropy 
evaluated at $T_c$ in \eqref{surengent}.

The important consequence of this qualitative difference is that, for sufficiently large values of $\kappa$ the phase transition 
between the ball and the ring configurations may disappear. As we can see from Fig.\ref{fig:pdgm}, such a phase transition does not 
exist for $\kappa = 0.5$. The critical value of $\kappa$ at which this phase transition 
ceases to exist is approximately $0.33 \approx 1/3$. Thus, we see that the temperature dependence of the surface tension can crucially affect the 
existence of the phase transitions between fluid configurations. In the dual gravity picture, this would have important consequence 
 for the phase transition between black holes of different horizon topologies. This calls for a future investigation, along the lines of 
\cite{Aharony:2005bm} from the gravity side to ascertain the value of $\kappa$.

Finally, we would like to observe that the parameters determining the validity of our analysis are the same as 
in \cite{Lahiri:2007ae}. The first of such parameters is given by the change in the fluid temperature over the scale of the 
mean free path $\Delta u \equiv \tilde \omega v_+ / (1-v_+^2)$, which must be small for the fluid approximation to be 
valid. The other parameter is $v_+/\tilde \omega$ for the ball and $\{v_- , v_+ - v_-\}/\tilde \omega$ for the rings,
which must be large for $\delta(f)$ to be well approximated by the Dirac delta. These parameters have a $\kappa$ dependence 
through $\tilde \omega$ for a fixed value of energy. Both these parameters are not significantly 
affected by the value of $\kappa$ (for the values of it that we have used), in the range 
of parameters that we consider. Therefore, we expect the validity of our result 
to be as good as  that in \cite{Lahiri:2007ae}.

%********************************************************************************
\section{Discussions}\label{sec:disc}
%********************************************************************************
%
In this paper, after performing a systematic analysis of the nature of surface transport 
in relativistic fluids, we were able to significantly constrain the structural form of the fluid equations 
at the surface. We have focused on some particular cases during our analysis, namely perfect 
fluids in arbitrary dimensions and the next to leading order corrections to 3+1 dimensional 
relativistic normal fluids. 

Although we have a specific set up at the back of our 
minds, as indicated in \S \ref{sec:intro}, our construction may be useful in more general settings,
like when boundaries between different fluid phases are present. Since we do not use any particular form 
for the distributions $\theta(f)$ and $\delta(f)$, they can be suitably chosen to model a 
wide variety of situations. In order to serve a more general purpose, it may be particularly useful
to study the non-relativistic surface effects. This may be achieved by taking 
a non-relativistic limit of our set up following \cite{Banerjee:2015hra, Banerjee:2015uta,Banerjee:2014mka}.

There are some immediate extensions of our work that are worth investigating.
For instance, it would be very interesting to work out the next to leading order surface effects in 
superfluids. Due to the interplay between the vector which is normal to the surface, 
and the superfluid velocity, there may be a very rich, but yet unexplored surface transport properties 
in this case. In fact, while analyzing the zeroth order superfluids in \S \ref{sec:superfl}, 
we noticed a new term, in the modified Young-Laplace equation \eqref{superYL}, which has not been 
widely considered in the literature. 

It would be very interesting to understand the implication of this term on the thermodynamics of 
finite lumps of superfluids. This may be accomplished by undertaking an analysis of various 
possible superfluid configurations along the lines discussed in \S \ref{sec:config}. 
In fact, such an analysis may also provide direct hints towards the existence of hairy black-rings or hairy black holes with 
other exotic horizon topologies in Scherk-Schwarz compactified $AdS$ spacetimes 
via the AdS/CFT correspondence. 

In section \S \ref{sec:config}, we have analyzed the effects of temperature dependence of 
surface tension on phase diagram of some simple fluid configurations. We found that 
this effect can be very significant, especially while drawing conclusions 
about the existence of a phase transition between the ball and ring type 
configurations. Since, we have considered our sample system to be the same 
as in \cite{Lahiri:2007ae}, our observation may have direct relevance for the 
existence of phase transition between spinning black holes and black-rings 
in Scherk-Schwarz compactified $AdS_5$. In particular, if the surface tension 
for these configurations scales as the inverse of temperature, then 
our analysis suggests that the existence of such a phase transition 
cannot be reliably predicted by a fluid dynamical analysis. This observation 
calls for generalization of \cite{Aharony:2005bm}, to deduce the exact dependence of
surface tension on temperature. 

Another interesting extension of our work is the possible generalization to embedded 
fluids with surfaces of higher co-dimension and its application to the description of asymptotically 
flat and AdS black holes. As it is well known, both Myers-Perry black holes and the higher-dimensional 
Kerr-AdS black holes admit ultraspinning regimes \cite{Emparan:2003sy, Armas:2010hz}. Moreover, it was shown in \cite{Emparan:2009vd, Armas:2010hz} that these regimes can be described by a 
rotating fluid disc with a boundary, where the fluid is moving at the speed of light. 
These analytic solutions, therefore, could allow us to extract some of the new surface transport coefficients 
that we have found in this work and hence study the physical and stability properties of these black holes using the description of fluid dynamics with surfaces.

% Along similar lines, it would also be interesting to push the analysis of 3+1 dimensional 
% charged fluids to one higher order (corresponding to second order in the bulk 
% and first order at the boundary). This may lead us to some very interesting 
% anomaly induced surface effects. The corresponding analysis for space filling fluids has 
% been performed in \cite{Bhattacharyya:2013ida}. A similar question, in the context 
% of fluid surfaces, generically for anomaly induced transport, 
% along the lines of developments in \cite{Jensen:2012kj, Jensen:2013kka, Jensen:2013rga}, 
% would also be very interesting to address. 
% 
% Another curious feature that we noticed in \S \ref{sec:chfirstord} was modification of 
% the Young-Laplace equation, by a term controlled by the $U(1)$ anomaly. This term 
% seems to augment the bulk pressure and create an anomaly induced force on the surface 
% of the fluid. It would be important to understand the physical meaning of this term more 
% clearly and here it may be helpful to study some concrete examples. 

We would like point out one curious feature related to anomaly induced transport properties 
(see \cite{Jensen:2012kj, Jensen:2013kka, Jensen:2013rga} for the most recent discussions on this). 
In our constructions here,  we have treated the bulk of the fluid by multiplying the partition 
function of space filling fluids with a $\theta(f)$ function. 
This procedure was justified (see \S \ref{sec:intro}) by noting that  $\theta(f)$ 
denoted the change in the bulk transport coefficients at or near the surface. 
Since the usual transport coefficients are macroscopic parameters, representing 
the microscopic UV theory only in an effective way, this procedure of introducing the 
$\theta(f)$ function is perfectly well-defined. However, there may exist certain terms 
in the partition function whose coefficients must be a constant as a consequence of gauge invariance \cite{Banerjee:2012iz}. 
The terms representing transport due to anomalies (for instance the term $\mathcal W_{\text{anom}}$ in \cite{Banerjee:2012iz}), 
also falls within a similar category, since their form is fixed by the criterion that they reproduce the right anomaly coefficient locally 
everywhere in spacetime, including at the surface of the fluid. Such an anomaly coefficient is 
not an effective macroscopic parameter but a parameter of the microscopic theory. Therefore, 
those terms cannot be straightforwardly handled in an effective way by multiplying the, with $\theta(f)$. 
\footnote{This is because it would then imply that the anomaly coefficient is varying in spacetime. Whether there 
can be a consistent microscopic theory where the anomaly coefficient can vary over spacetime may be an interesting question in itself.} We postpone further analysis of these terms, in the context of 
fluid surfaces, to future work.

It would be interesting to see how the transport coefficients discussed in this paper fit into the 
classification of \cite{Haehl:2015pja, Haehl:2014zda}. 
% Although a more thorough comparisn may be required to eliminate 
% any possible subtleties. 
Further, recently there has been significant progress in formulating dissipative fluid 
dynamics in terms of an action \cite{Crossley:2015evo, Haehl:2015uoc}. It would be very interesting to 
understand how the presence of surfaces generalizes these constructions. 
In fact, it would be particularly interesting to understand 
the time evolution of fluid surfaces, involving dissipation. 
If we are able to incorporate time dependence, in a controlled 
fashion within our set up, it may have some relevance to situations concerning dynamical 
formation of surfaces, in the interface of two phases, which are described by the 
Cahn-Hilliard equations \cite{1958JChPh..28..258C}.

%
%~~~~~~~~~~~~~~~~~~~~~~~~~~~~~~~~~~~~~~~~~~~~~~
\acknowledgments 
%~~~~~~~~~~~~~~~~~~~~~~~~~~~~~~~~~~~~~~~~~~~~~~
We would like to thank Diptarka Das for many useful discussions and for collaboration at the initial stages of the project. 
We are also especially thankful to Pallab Basu and Sayantani Bhattacharyya  for many useful and insightful discussions. We would also like to thank 
Nabamita Banerjee, Shamik Banerjee, Buddhapriya Chakrabarti, Sitikantha Das, Aristomenis Donos, Suvankar Dutta, Felix Haehl, Carlos Hoyos, Veronika Hubeny, 
Akash Jain, Sachin Jain, Dileep Jatkar, Juan Jottar, Sandipan Kundu,  Wei Li, R. Loganayagam, Shiraz Minwalla, Andrew O'bannon, 
Mukund Rangamani, Samriddhi Sankar Ray, Simon Ross and Tarun Sharma for several useful discussions. 
We are gratful to Felix Haehl, Simon Ross and Tadashi Takayanagi for valuable comments on the draft of this manuscript. 
JB and NK would like to thank IITK for hospitality, where a part of this project was completed. JA would like to thank NBI for hospitality during 
the course of this project. JA acknowledges the current support of the ERC Starting Grant 335146 HoloBHC. 
JB is supported by the STFC Consolidated Grant ST/L000407/1. 
% 
% %****************************************
 \appendix
% %***************************************
% 
%**********************************************************************************************************
%This appendix is added in v2
% %*******************************************************************************
 \section{Frame transformation in the bulk}\label{app:Frametrans}
% %*******************************************************************************

In this appendix, we shall perform a frame transformation from the Landau-frame in the bulk of the 
fluid, to the orthogonal-Landau-frame which was defined in \S\ref{sssec:genpart2}. 

In the presence of the surface at $f(x)=0$, we can choose our coordinates 
so that one of the spatial coordinates vanishes at the surface. Let us refer to such a coordinate by $f$. 
For sufficiently well behaved spacetimes, the constant $f$ surfaces would foliate the 
entire spacetime, including the bulk of the fluid. Every point on the constant $f$ surfaces 
would admit a well defined, outward pointing normal vector, which we refer to as $n_\mu$. This provides us with an 
extension of the normal vector on the surface throughout the spacetime.
\footnote{This extension of $n_{\mu}$ is clearly non-unique. But in our description, 
this ambiguity is absorbed into the ambiguity related to choice of frames for the bulk fluid variables.}

As discussed in \S \ref{sssec:genpart2}, instead of imposing the Landau-frame condition 
\begin{equation}\label{Lfrm}
 u^{\mu} T_{\mu \nu} = -\mathcal E u_{\nu}
\end{equation}
in the bulk, we make a slightly different frame choice, which is given by 
\begin{equation}\label{OLfrm}
  \mathcal H^{\alpha \nu} \tilde u^{\mu} T_{\mu \nu} = -\mathcal E \tilde u^{\alpha}, ~~\tilde u \cdot n = 0 
\end{equation}
to all orders, everywhere in the bulk of the fluid. Here $\mathcal H_{\mu \nu} = \mathcal G_{\mu \nu} - n_\mu n_\nu$, is the 
projector orthogonal $n_\mu$, which is defined throughout the bulk of the fluid. It possible to impose this condition everywhere in the bulk, 
since we now have a definition of $n_\mu$ extended throughout the bulk of the fluid. This immediately ensures that the fluid velocity 
is orthogonal to the normal vector at the surface of the fluid, where $n_\mu$ is unambiguously defined. 
This frame transformation may be achieved by simply redefining 
\begin{equation}\label{frmtransform}
 \tilde u^{\mu} = u^{\mu} - n^{\mu} \left(u\cdot n \right)
\end{equation}
Now, at the leading order, for stationary configurations, the fluid velocity can be oriented along the time-like killing vector, preserving 
the Landau-frame condition. 
Since, the surface has a trivial time evolution for stationarity configurations, this immediately implies $n\cdot u^{(0)}$ must be zero. 
This is no longer true at higher orders and we need to perform a frame transformation by higher derivative terms in order to 
achieve \eqref{OLfrm}.

In the partition function construction presented in \cite{Banerjee:2012iz}, the fluid velocity and temperature in the bulk, are solved in terms of the 
background fields. The first non-trivial corrections to the fluid velocity occurs at the second order in derivatives. This implies that we have to 
perform a frame transformations with second order terms, and the transformation should have a form like \eqref{frmtransform}. In fact, the exact form 
of the required transformation can be read off from the second order corrections to the fluid velocity in \cite{Banerjee:2012iz}
\footnote{Note that the combination of the second order terms are chosen such that in the stationary situation, when expressed in terms 
of the background data, it reduces to the velocity corrections obtained in \cite{Banerjee:2012iz}.  
This choice may not be unique, specially when applying this trick to arbitrary orders, but as long as we focus only on stationary 
configurations, all such frames would be equivalent.}
\begin{equation}\label{frmtransfull}
   u^{\mu} = \tilde u^{\mu} + n^{\mu} \left( \left( \tilde v_1 ~\mathfrak a_{\alpha} \omega^{\alpha \nu} 
  +  \tilde v_2  ~P^{\nu}_{\alpha} \nabla_{\beta} \omega^{\beta \alpha} \right) n_\nu \right) + \dots,
\end{equation}
where $P_{\mu \nu} = \mathcal G_{\mu \nu} + u_{\mu} u _{\nu}$, is the projector orthogonal to the fluid velocity and the ellipsis denote the higher 
order order corrections that may be necessary to keep \eqref{OLfrm} intact. We should take the coefficients $\tilde v_1$ and $\tilde v_2$ to be the same 
as the ones appearing in the second order velocity corrections, worked out in \cite{Banerjee:2012iz}.

This frame transformations directly impacts the form of the second order 
stress tensor. In the Landau-frame the second order bulk stress tensor was given by  \eqref{STLandau}, and is now modified to 
\begin{equation}\label{STOrthLandau}
\begin{split}
T_{\mu \nu} =& T \left( \ \kappa_1 ~\tilde R_{\langle\mu \nu\rangle} + \kappa_2 ~K_{\langle\mu \nu\rangle} + \lambda_3  \ 
\omega_{\langle\mu}^{\,\, \alpha} \omega_{\alpha \nu\rangle} + \lambda_4 \ \mathfrak{a}_{\langle \mu}\mathfrak{a}_{\nu \rangle} 
 + P_{\mu \nu}(\zeta_2  \tilde R+\zeta_3  \tilde R_{\alpha \beta}u^\alpha u^\beta +\xi_3 \omega^2+\xi_4 \mathfrak{a}^2 ) \ \right)~~\\
 &+\left(\tilde u_{( \mu} n_{\nu )}  \left( v_1 ~\mathfrak a_{\alpha} \omega^{\alpha \rho} n_{\rho}
  +  v_2  ~P^{\rho}_{\alpha} \nabla_{\beta} \omega^{\beta \alpha} n_{\rho} \right) \right)~~.
\end{split}
\end{equation}
Here $v_m = 2 (\mathcal E + P) \tilde v_m$.
All the second order fluid quantities in \eqref{STOrthLandau} are to be expressed in terms of the transformed velocity $\tilde u_{\mu}$, 
and the vector field $n_\mu$ is the extension of the normal vector at the surface throughout the bulk, as explained above. 

Now, we know from the partition function analysis that there are only 3 independent bulk transport coefficients and therefore 
$v_1$ and $v_2$ must be related to the rest of the transport coefficients, through two new relations. 
The necessary frame transformation to ensure \eqref{OLfrm} fixes them to be 
\begin{equation}
 v_1 = -T^2 \partial_T \left(2 \kappa_1 + \kappa_2 - \lambda_3  \right), 
 v_2 = T \left( 2 \kappa_1 + \kappa_2 - \lambda_3  \right)~~.
\end{equation}
These relations must hold in addition to the five relation between the rest of the transport coefficients in \eqref{STOrthLandau} as explicated 
in \cite{Banerjee:2012iz}. The fluid velocity $\tilde u_\mu$ that will now be obtained in terms of the background data, 
when we compare \eqref{STOrthLandau} with the bulk partition 
function in \eqref{sum:surpartsecord}, will be automatically projected orthogonal to $n_\mu$, 
which has been ensured due to the frame choice \eqref{OLfrm}. 

%**********************************************************************************************************

%***************************************************************************************************************************
\section{General constraints on the stress tensor}\label{ssec:aaa}
%***************************************************************************************************************************
In this appendix we discuss generic constrains and symmetries of the surface stress tensor \eqref{surPhenoST}. 
The full spacetime stress tensor, including the bulk contribution, to second order in derivatives, can be 
decomposed as in \eqref{STschC}, where the surface stress tensor \eqref{surPhenoST} is written in the form
\begin{equation} \label{eq:gstress1}
T^{\mu\nu}_{\text{surf}}=T^{\mu\nu}_{(1)}\tilde\delta(f)+T^{\mu\nu\rho}_{(2)}\partial_\rho \tilde\delta(f)~~.
\end{equation}
Here, the structure $T^{\mu\nu}_{(1)}$ denotes the contribution to the surface stress tensor of a monopole source of stress while $T^{\mu\nu\rho}_{(2)}$ denotes the contribution of a dipole source of stress. When applying this decomposition to \eqref{surPhenoST} we easily read off
\begin{equation} \label{eq:gstress2}
T^{\mu\nu\rho}_{(2)}=\left(\mathfrak{s1}_{4}\mathcal{P}^{\mu\nu}+\mathfrak{s2}_{4}u^{\mu} u^{\nu}+\mathfrak{s3}_{4}n^{\mu}n^{\nu}+\mathfrak{s4}_{4}u^{(\mu}n^{\nu)}\right)n^{\rho}+\mathfrak{v1}_{7}u^{(\mu}\mathcal{P}^{\nu)\rho} +\mathfrak{v2}_{7}n^{(\mu}\mathcal{P}^{\nu)\rho}~~,
\end{equation}
while $T^{\mu\nu}_{(1)}$ includes all the other surface stress tensor components. 

However, we can impose additional constraints which follows from the fact that the 
the stress tensor \eqref{eq:gstress1} enjoys a 
symmetry, for which its components transform as (see \cite{Vasilic:2007wp} for more details)
\begin{equation}\label{Ttrans}
\delta T^{\mu\nu}_{(1)}=-\epsilon^{\mu\nu a}v_a~~,~~\delta T^{\mu\nu\rho}_{(2)}=\epsilon^{\mu\nu a}e_a^{\rho}~~,
\end{equation}
for some coefficients $\epsilon^{\mu\nu a}$ and where we recall that we have defined $v_a={e_{a}}^{\rho} n^{\mu}\nabla_\mu n_\rho$. This transformation arises due to the freedom of introducing $(D-1)$ redundant delta functions in \eqref{eq:gstress1}, so that \eqref{eq:gstress1} could have been written as
\begin{equation}
T^{\mu\nu}_{\text{surf}}=\int_{\partial \mathcal{M}}\left(T^{\mu\nu}_{(1)}\tilde\delta^{(D)}(f)+T^{\mu\nu\rho}_{(2)}\partial_\rho \tilde\delta^{(D)}(f)\right)~~,
\end{equation}
where $\tilde \delta^{(D)}(f)=\sqrt{\partial f . \partial f}\delta^{(1)}(x^{1})\delta^{(2)}(x^{2})...\delta^{(D-1)}(x^{D-1})\delta (f)$. 
Therefore, the tangential derivatives of the distribution $\tilde\delta(f)$ are integrated out and 
the coefficients $T^{\mu\nu a}_{(2)}$ can be removed \cite{Vasilic:2007wp}. 
This implies that the terms involving $\mathfrak{v1}_{7}$ and $\mathfrak{v2}_{7}$ in \eqref{eq:gstress2} 
can be set to zero.\footnote{In this paper we assume that the surface does not have boundaries. 
However, if there were boundaries then this symmetry would not present at the surface boundary 
and there we would need to impose $\epsilon^{\mu\nu a}\eta_a=0|_{\partial \mathcal{M}}$ for a normal covector $\eta_a$ to the surface boundary.} 
However, the stress tensor that follows from the partition function is obtained in a fixed gauge, as far as the transformations 
\eqref{Ttrans} are concerned. In that case, although we do get $\mathfrak{v1}_{7} = 0$, $\mathfrak{v2}_{7}$ however, 
is related to other transport coefficients (see \eqref{mainconst}). 

Furthermore, for a stress tensor of the form \eqref{STschC}, there is a perturbative symmetry 
that allows to displace the surface located $f=0$ by a small amount $\varepsilon$ such 
that $f\to f+\varepsilon$. This symmetry expresses the freedom of defining the surface theory on a specific infinitely thin slice of a surface with finite thickness. Looking at Fig.~\ref{fig:thetadelta}, this means slightly displacing the dashed vertical line into another location within the distribution $\tilde\delta(f)$.  Under this infinitesimal displacement, the form of the stress tensor \eqref{STschC} is unchanged but its components have varied according to
\begin{equation} \label{eq:sym2}
\delta T^{\mu\nu}=\varepsilon\left(-T^{\mu\nu}_{(0)}+n^{\rho}\partial_\rho T^{\mu\nu}_{(1)}\right) \tilde \delta (f)+\varepsilon T^{\mu\nu}_{(1)}n^{\rho}\partial_\rho \tilde\delta(f)~~.
\end{equation}
From \eqref{eq:sym2} we see that this transformation induces a contribution proportional to $n^{\rho}\partial_\rho \tilde\delta(f)$. If we take $T^{\mu\nu}_{(1)}$ to have the perfect fluid form at leading order, then by 
appropriately choosing $\varepsilon$ we could work with a surface for which either $\mathfrak{s1}_{4}$ or $\mathfrak{s2}_{4}$ vanish. 
However, since the transformation \eqref{eq:sym2} induces a term proportional to $T^{\mu\nu}_{(0)}\tilde\delta(f)$, such 
choice of surface would require to introduce the bulk pressure $P|_{f=0}$ as an independent scalar in the surface 
part of the partition function. For this reason we have decided to work with the 
scalars $K$ and $\mathfrak{a}^{\mu}n_\mu$ instead.\footnote{When the bulk pressure vanishes at the surface, 
which is the case studied in \cite{Armas:2013goa}, then $K$ and $\mathfrak{a}^{\mu}n_\mu$ are not independent.}

There are structural consistency conditions on the components of the surface stress tensor \eqref{eq:gstress1}, which, 
therefore, are not all independent and thus cannot be freely chosen. 
These consistency conditions arise due to the fact that we are working 
with the expansion \eqref{eq:gstress1} to a particular order and they can 
be derived by carefully analysing the conservation equation  \eqref{eq:conVV} 
to this particular order. 
The resulting conditions must hold in any physical situation, including time-dependent settings. One of these conditions constrains the dipole source of stress such that \cite{Vasilic:2007wp}
\begin{equation}
n_{\mu}n_{\nu}n_{\rho}T^{\mu\nu\rho}_{(2)}=0~~,
\end{equation}
and also, for codimension-1 surfaces, it follows that we must have
\begin{equation}
{e^{a}}_{\mu}n_\nu n_\rho T^{\mu\nu\rho}_{(2)}=0~~,
\end{equation}
which is a trivial consequence of there being no transverse two-plane on which the surface can rotate \cite{Armas:2013hsa}. In turn, both conditions imply the constraints
\begin{equation} \label{eq:c1}
\mathfrak{s3}_4=0~~,~~\mathfrak{s4}_4=0~~.
\end{equation}

The remaining conditions determine the normal components of the monopole source $T^{\mu\nu}_{(1)}$ in terms of the dipole source of stress $T^{\mu\nu\rho}_{(2)}$. In particular we must have that
\begin{equation}
\begin{split} \label{eq:cond1}
n_\mu n_\nu T^{\mu\nu}_{(1)}&=\left(T^{ab\rho}_{(2)} K_{ab}n_\rho-n_{\lambda}n_\rho T^{\lambda\rho a}_{(2)}v_a\right) \\
&=\left(\mathfrak{s1}_4K+(\mathfrak{s1}_4+\mathfrak{s2}_4)\mathfrak{a}^{\rho}n_\rho\right)~~.
\end{split}
\end{equation}
If we now evaluate the normal components of $T^{\mu\nu}_{(1)}$ in \eqref{surPhenoST} we obtain
\begin{equation}
n_\mu n_\nu T^{\mu\nu}_{(1)}=\left(\mathfrak{s3}_1\mathfrak{a}^{\rho}n_\rho+\mathfrak{s3}_2K+\mathfrak{s3}_3\ell^{\rho}n_\rho\right)~~,
\end{equation}
which upon comparison with \eqref{eq:cond1} leads to the constraints
\begin{equation} \label{eq:c2}
\mathfrak{s3}_1=-(\mathfrak{s1}_4+\mathfrak{s2}_4)~~,~~\mathfrak{s3}_2=\mathfrak{s1}_4~~,~~\mathfrak{s3}_3=0~~.
\end{equation}
Moreover, the remaining normal components of $T^{\mu\nu}_{(0)\text{surf}}$ must respect the following condition
\begin{equation}\label{eq:cond2}
\begin{split}
{e^{a}}_{\mu} n_\nu T^{\mu\nu}_{(1)}&=-2 n_\rho\left( \nabla_b T^{ab\rho}_{(2)}+\left( T^{ac\rho}_{(2)}+T^{a\rho c}_{(2)}\right)v_c\right)~~\\
&=-2{\mathcal{P}^{a}}_{b}\left(\left(-T\partial_T\mathfrak{s1}_4+\mathfrak{s1}_4+\mathfrak{s2}_4\right)\mathfrak{a}^{b}+\left(\mathfrak{s1}_4+\frac{1}{2}\mathfrak{v2}_7\right)v^{b}\right)~~.
\end{split}
\end{equation}
Performing the same operation in \eqref{surPhenoST} yields
\begin{equation}
{e^{a}}_{\mu} n_\nu T^{\mu\nu}_{(1)}=\sum_{i=1}^{6}\mathfrak{v2}_{i}\mathcal{V}^{a}_{(i)}+\frac{1}{2}\sum_{i=1}^{3}\mathfrak{s4}_{i}\mathcal{S}_{(i)}u^{a}~~,
\end{equation}
which, again, upon comparison with \eqref{eq:cond2} leads to the constraints
\begin{equation} \label{eq:c3}
\begin{split}
&\mathfrak{v2}_{i}=0~\forall~i~\in~\{2,4,5,6\}~~,~~\mathfrak{v2}_{1}=2\left(T\partial_T\mathfrak{s1}_{4}-\mathfrak{s1}_{4}-\mathfrak{s2}_{4}\right)~~,~~\mathfrak{v2}_{3}=-2\left(\mathfrak{s1}_{4}+\frac{1}{2}\mathfrak{v2}_{7}\right)~~, \\
&\mathfrak{s4}_{i}=0~\forall~i~\in~\{1,2,3\}~~.
\end{split}
\end{equation}
As it can be quickly verified, the constraints \eqref{eq:c1},\eqref{eq:c2} and \eqref{eq:c3} are a subset of the contraints captured by the partition function analysis \eqref{mainconst}. From these considerations, we see that the only components which are left unconstrained due to stress tensor conservation are the surface components of the stress tensor $T^{ab}_{(1)}$ and the dipole source components $T^{ab\rho}_{(2)}$, also known as the bending moment. The remaining constraints in \eqref{mainconst} in the case of relativistic fluids can be obtained by demanding positivity of the entropy current, an analysis which is carried in Appendix~\ref{sec:entropy}.

%
%********************************************************************************
\section{Entropy current constraints} \label{sec:entropy}
%******************************************************************************** 
In this appendix we analyze the constraints on the transport coefficients that arise from the positivity of the entropy current and show that both the partition function and the effective action capture these constraints. This can be done by analyzing the divergence of the entropy current for a membrane subjected to external forces. The equations of motion were given in \eqref{eq:conservation} and\eqref{eq:mYL}.
The fact that the equations of motion only involve $\mathbb{T}^{ab}_{\text{sur}}$, $T^{ab\rho}_{(2)}$ and $T^{\mu\nu}_{(0)}$, signifies that only these three structures are required in order specific the dynamics of the membrane. However, we need to specify what the conservation equation for the surface entropy current is. The full entropy current can be expanded analogously to the stress tensor,
\begin{equation} \label{eq:entropy}
J^{\mu}_{s}=J^{\mu}_{s(0)}\theta(f)+J^{\mu}_{s(1)}\tilde\delta(f)+J^{\mu\rho}_{s(2)}\partial_{\rho}\tilde\delta(f)+...~~,
\end{equation}
where the surface part, up to first order, is given by
\begin{equation}
J^{\mu}_{s~sur}=J^{\mu}_{s(1)}\tilde\delta(f)+J^{\mu\rho}_{s(2)}\partial_{\rho}\tilde\delta(f)~~.
\end{equation}
For stationary configurations, we require \eqref{eq:entropy} to be divergence free. In the bulk, this simply results in the bulk conservation equation $\nabla_\mu J^{\mu}_{s(0)}=0$ while in the surface this results in
\begin{equation} \label{eq:entropyc}
\nabla_a\mathcal{J}^{a}_{s}=J^{\mu}_{s(0)}n_\mu|_{f=0}~~,
\end{equation}
where \cite{Armas:2013aka}
\begin{equation}
\mathcal{J}^{a}_{s}=J^{a}_{s(1)}-{e^{a}}_{\mu}\nabla_\rho J^{\mu\rho}_{s(2)}-{e^{a}}_{\mu}\nabla_a J^{\mu b}_{s(2)}+n_\rho J^{b\rho}_{s(2)}{K^{a}}_b~~.
\end{equation}
Here we have assumed that the entropy current can be obtained from the partition function/action in a similar way as a $U(1)$ charge current, in the spirit of \cite{Haehl:2015pja}. From the effective action \eqref{eq:action}, the surface entropy current can be obtained via the variation \cite{Armas:2013goa, Armas:2014rva}
\begin{equation} \label{eq:newc}
\mathcal{J}^{a}_{s}=\frac{\partial \mathcal{I}_{\text{surf}}}{\partial T}u^{a}~~,
\end{equation}
though, depending on the type of corrections that the bulk action receives, there may be contributions to $\mathcal{J}^{a}_{s}$ due to bulk terms. Note that we have assumed that it is always possible to write the entropy current that follows from an action in the form \eqref{eq:newc}, which does not include all terms allowed by symmetry. The reason for this is that for stationary configurations we are always free to add total derivative terms to the entropy current, which are also divergence free, such that it takes the form \eqref{eq:newc}. This is true for uncharged fluids up to second order in derivatives, both for the bulk entropy current and for the surface entropy current. Since \eqref{eq:entropyc} depends only on $\mathcal{J}^{a}_{s}$ and $J^{\mu}_{s(0)}$ it is only necessary to classify these terms in order to obtain the constraints on the fluid transport. 

Before proceeding and classifying possible terms that can appear in the different relevant structures, it is important to properly define fluid frames both in the bulk and in the boundary. Because, in principle, we can be placing a completely different fluid on the surface of another bulk fluid, we should consider two fluids, each described by their own fluid variables. In this appendix, we use \emph{tilde} quantities to describe the bulk fluid, which is characterized by the set of bulk variables $(\tilde T, \tilde u^{\mu})$, while the surface fluid is characterized by the set of surface variables $(T, u^{\mu})$. However, in order to fully specify the system, we need to impose boundary conditions on the bulk fluid variables. These boundary conditions were described in \eqref{eq:boundaryc} and a natural consequence of them is that
\begin{equation} \label{eq:boundaryc1}
\tilde u^{\mu}\nabla_\mu \tilde T|_{f=0}= u^{a}\nabla_a T~~.
\end{equation}
These boundary conditions are dynamical, in the sense that the evolution of $(\tilde T, \tilde u^{\mu})$, described by the bulk equations to leading order
\begin{equation} \label{eq:bulkv}
\tilde u^{\mu}\nabla_\mu \tilde T=-\tilde s\frac{\partial \tilde T}{\partial\tilde s}\tilde\Theta~~,~~P^{\mu\nu}\nabla_\nu\tilde T=-\tilde T \tilde{\mathfrak{a}}^{\mu}~~,
\end{equation}
where $\tilde \Theta=\nabla_\mu\tilde u^{\mu}$, must be subjected to the boundary conditions \eqref{eq:boundaryc}, which are dynamically determined by the surface evolution equations to leading order
\begin{equation} \label{eq:surfv}
 u^{a}\nabla_a T=-s\frac{\partial  T}{\partial s}\Theta~~,~~\mathcal{P}^{ab}\nabla_a T=- T {\mathfrak{a}}^{b}~~.
\end{equation}
Note that \eqref{eq:boundaryc1} states that derivatives of the temperature along the fluid flows are equal in the bulk and in the surface. Derivatives of the temperature tangentially to the surface, but perpendicular to the fluid flows, are also guaranteed by \eqref{eq:boundaryc} to be the same in both the bulk and in the surface. This can be seen by tangentially projecting the second equation in \eqref{eq:bulkv} and comparing it with the second equation in \eqref{eq:surfv}. In order to more clearly present our results, we will impose \eqref{eq:boundaryc} from the get go while carefully keeping track of the derivatives of $\tilde u^{\mu}$ and $\tilde T$ using equations \eqref{eq:bulkv}. 

We now make a few comments regarding fluid frame transformations of the bulk and surface fluids. Frame transformations of the bulk fluid variables $(\tilde T, \tilde u^{\mu})$ 
allows us to set the bulk stress tensor in the orthogonal-Landau frame \eqref{OLfrm}, that is, 
\begin{equation} \label{eq:Landau}
\mathcal T^{\mu\nu}_{(0)}\tilde u_{\mu}{\mathcal{H}^{\lambda}}_\nu = 0~~,
\end{equation}
where $\mathcal T^{\mu\nu}_{(0)}$ are the higher derivative corrections to $T^{\mu\nu}_{(0)}$. 
However, due to the boundary condition \eqref{eq:boundaryc}, the restriction of such 
frame transformations to the surface at $f=0$ will induce a 
frame transformation of the surface fluid variables $(T, u^{\mu})$, if they are defined in the Landau frame.
However, since we are working with stationary fluids up 
to second order in bulk derivatives, then such bulk frame 
transformations are second order in derivatives. Since the surface 
fluid quantities are only expanded to first order, then bulk frame transformations 
do not affect the surface stress tensor neither the surface entropy current. 
On the other hand, we can perform a first order frame transformation of the surface fluid 
variables $(T, u^{\mu})$ and set the surface stress tensor in the Landau frame
\begin{equation} \label{eq:Landau1}
\mathbb{{T}_{(j)}}^{ab}_{\text{sur}}u_{b}= 0 ~~,
\end{equation} 
where again, $\mathbb{{T}_{(j)}}^{ab}_{\text{sur}}$ are the higher derivative corrections to $\mathbb{{T}}^{ab}_{\text{sur}}$.

This is not an elegant choice, in the sense that if we impose \eqref{eq:Landau1} for the surface stress tensor then we cannot simultaneously impose \eqref{eq:Landau} for the bulk stress tensor evaluated at the surface. However, this is still a convenient choice because, as it will be explained below, the analysis of the divergence of the entropy current is insensitive to such frame transformations of the bulk stress tensor at the surface besides the fact that it reduces the number of structures appearing in the stress tensor to 4. We note, furthermore, that frame transformations do not modify the components $T^{ab\rho}_{(2)}$ of the stress tensor \cite{Armas:2013goa}.

We now proceed and write the relevant terms that enter the several structures involved. We note that, since we are working in the orthogonal-Landau frame \eqref{eq:Landau}-\eqref{eq:Landau1} and since the entropy conservation equation \eqref{eq:entropyc} only involves the projection of Eq.~\ref{eq:conservation} along $u_{\mu}$, we find, using \eqref{STOrthLandau}, that
\begin{equation} \label{eq:force}
T^{\mu\nu}_{(0)}u_{\nu}n_\mu|_{f=0}=-\frac{1}{2}\left( v_1 ~\tilde{\mathfrak{a}}_{\alpha} \tilde \omega^{\alpha \nu} n_{\nu}+  v_2  ~P^{\nu}_{\alpha} \nabla_{\beta} \tilde \omega^{\beta \alpha} n_{\nu} \right) ~~.
\end{equation}
Moreover, the fact that only the contraction matters $T^{\mu\nu}_{(0)}u_{\nu}n_\mu|_{f=0}$ for the divergence of the entropy current implies that surface fluid frame transformations do not affect it.  On the other hand, the classification of $J^{\mu}_{s(0)}$ has already been done in \cite{Bhattacharyya:2012nq}. As we will see later in this section, when comparing with the partition function and effective action, the contraction $J^{\mu}_{s(0)}n_\mu|_{f=0}$ for stationary configurations can be written as
\begin{equation} \label{eq:genentropy}
\begin{split}
J^{\mu}_{s(0)}n_\mu|_{f=0}=&~\pi_1 \Theta \mathfrak{a}^{\mu}n_\mu+\pi_2\tilde{\omega}^{\mu b}n_\mu \mathfrak{a}_b+\pi_3u^{a}\nabla_a\left(\mathfrak{a}^{\mu}n_\mu\right) \\
&+\pi_4 u^{a}\mathfrak{a}^b{K_{ab}}+\pi_5\nabla_a\left(\tilde\omega^{\mu a}n_\mu\right)+\pi_6u^{a}\nabla_a K+\pi_7u^{a}\nabla_b{K^{b}}_a~~\\
&+\tilde s\left( v_1 ~\tilde{\mathfrak{a}}_{\alpha} \tilde \omega^{\alpha \nu} n_{\nu}+  v_2  ~P^{\nu}_{\alpha} \nabla_{\beta} \tilde \omega^{\beta \alpha} n_{\nu} \right)~~,
\end{split}
\end{equation}
where the last line above is due to the frame change \eqref{frmtransfull}. Here the transport coefficients $\pi_i$ are only functions of $T$ since we have restricted it to the surface. As the analysis of \cite{Bhattacharyya:2012nq} shown, in a generic situation only 5 of the $\pi_i$ coefficients are independent but when stationarity is imposed, only 3 are independent \cite{Banerjee:2012iz}. Furthermore, the analysis of the remaining structures has largely been done in \cite{Armas:2013goa} but because of the presence of the bulk fluid and parity odd transport, we have
\begin{equation} \label{eq:Tab}
\mathbb{T}^{ab}_{\text{sur}}=\chi\mathcal{H}^{ab}+(\chi+\chi_E) u^{a}u^{b}+\alpha_1 K \mathcal{P}^{ab}+\alpha_2 \mathfrak{a}^{\mu}n_\mu \mathcal{P}^{ab}+\alpha_3 \ell^{\mu}n_\mu \mathcal{P}^{ab}+\alpha_4\mathcal{P}^{a}_c\mathcal{P}^{b}_dK^{cd}~~,
\end{equation}
\begin{equation}
T^{ab\rho}_{(2)}=\vartheta_1 \mathcal{H}^{ab}n^{\rho}+\vartheta_2 u^{a}u^{b}n^{\rho}~~,
\end{equation}
\begin{equation} \label{eq:s_surf}
\mathcal{J}^{a}_{s}=su^{a}+\gamma_1 Ku^{a}+\gamma_2 \mathfrak{a}^{\mu}n_\mu u^{a}+\gamma_3 \ell^{\mu}n_\mu u^{a}+\gamma_4 u^{b}{K_{b}}^{a}+\gamma_5 \tilde \omega^{\mu a}n_\mu +\gamma_6\epsilon^{abc}u_b\mathfrak{a}_c~~.
\end{equation}
It is worthwhile keeping in mind that all the 12 surface transport coefficients are only functions of $T$.\footnote{We could have added other terms to $\mathcal{J}^{a}_{s}$ such as terms proportional to $\mathfrak{a}^{a}$ and $v^{a}$, however, these terms would be required to vanish at the end and hence, for simplicity, we have not considered them.} Comparing with the work in \cite{Armas:2013goa} for a free membrane, we note that the terms $\alpha_2,\alpha_3,\vartheta_2,\gamma_3,\gamma_5,\gamma_6$ were not present. As mentioned in Sec.~\ref{ssec:aaa}, when the bulk pressure $P$ vanishes at $f=0$, then according to \eqref{genLY}, the scalars $K$ and $\mathfrak{a}^{\mu}n_\mu$ are not independent and hence we need to include $\alpha_2$ and $\gamma_2$. Consequently, we must also include the coefficient $\vartheta_2$. Removing it would require, as discussed in Sec.~\ref{ssec:aaa}, 
considering an extra term in $\mathbb{T}^{ab}_{\text{sur}}$ of the form $P|_{f=0}\mathcal{P}^{ab}$. 
The term $\gamma_6 \tilde \omega^{\mu a}n_\mu$ is a consequence of the presence of the bulk degrees of freedom.
For stationary configurations, it may be replaced by a term of the form $\mathcal{P}^{a}_c u^{b}{K_{b}}^{c}$
at the leading order; however, we must include it as a separate terms since 
they differ at higher orders.
\footnote{ We do not consider terms which vanish  for stationary configurations, upto the order we are keep tract of. 
In principle, such terms should be considered as well but it is possible to show that they do not contribute to the analysis of 
the entropy current of stationary configurations.} 
Finally, the coefficients $\alpha_3,\gamma_3,\gamma_6$ are well known in the context of parity odd fluids in 2+1 dimensions \cite{Jensen:2011xb} and, as noticed in \cite{Jensen:2011xb}, $\gamma_3$ and $\gamma_6$ are not independent. Due to the freedom of adding to \eqref{eq:s_surf} a total derivative term of the form $\nabla_b\left(\tilde\alpha\epsilon^{abc}u_c\right)$ for an arbitrary $\tilde\alpha$, shifting the coefficients $\gamma_3\to\gamma_3-\tilde \alpha$ and $\gamma_6\to \gamma_6+T\partial_T \tilde \alpha -\tilde \alpha$, then only the linear combination 
\begin{equation} \label{eq:linear}
\gamma_7=T\frac{\partial \gamma_3}{\partial T}+\gamma_6-\gamma_3~~,
\end{equation}
is invariant under this shift. The surprisingly simple form of \eqref{eq:Tab} can be obtained from \eqref{surPhenoST} by bringing it to the Landau frame. 

Given these structures, we now impose the entropy conservation equation \eqref{eq:entropyc} and, using \eqref{eq:conservation}, obtain an expression of the following form
\begin{equation} \label{eq:divergence}
\begin{split}
\nabla_a\mathcal{J}^{a}_{s}-J^{\mu}_{s(0)}n_\mu|_{f=0}=&~\beta_1 \Theta K+\beta_2 \Theta \mathfrak{a}^{\mu}n_\mu+\beta_3u^{b}\nabla_a {K^a}_b+\beta_4 u^{a}\mathfrak{a}^{b}K_{ab}+\beta_5u^{a}\nabla_a K \\
&+\beta_6u^{a}\nabla_a\left(\mathfrak{a}^{\mu}n_\mu\right)+\beta_7\sigma^{ab}K_{ab}+\beta_8\tilde\omega^{\mu b}n_\mu\mathfrak{a}_b+\beta_9\nabla_b\left(\tilde\omega^{\mu b}n_\mu\right) \\
&+\beta_{10} \Theta \ell^{\mu}n_\mu ~~.
\end{split}
\end{equation}
We note here that the effect of the last line of \eqref{eq:genentropy} is cancelled by the force term \eqref{eq:force}. Close inspection of \eqref{eq:divergence} leads us to conclude that all terms involved are linear in fluid data. Since we must set the RHS of \eqref{eq:divergence} to zero, according to \eqref{eq:entropyc}, then all $\beta_i$ coefficients must individually vanish, i.e., we must require that $\beta_i=0~,~\forall~i=1,...,10$. We can solve these constraints in terms of $\vartheta_i,\gamma_i$ and the external force coefficients $\pi_i$ leading to
\begin{equation} \label{eq:relations}
\begin{split}
&\alpha_4=\gamma_4 T~~,~~\gamma_5=\pi_5~~,~~\gamma_2 T=\pi_3+\vartheta_2~~,~~\gamma_1 T=\pi_6T-\vartheta_1~~,~~\gamma_4T=\pi_7T+2\vartheta_1~~,\\
&\alpha_1T^{-1}=\gamma_1-\frac{\partial \gamma_1}{\partial T} s\frac{\partial T}{\partial s}~~,~~2\frac{\partial\vartheta_1}{\partial  T}=\pi_4+\gamma_4+\frac{\partial\gamma_4}{\partial T}T~~,\\
&\frac{\partial\gamma_5}{\partial  T} T=-\pi_2~~,~~\alpha_3 T^{-1}=-\gamma_7T\frac{\partial T}{\partial s}~~,\\
&{\alpha_2}T^{-1}=-\pi_1-\left(2\frac{\partial (\vartheta_1-\vartheta_2)}{T\partial T} +\frac{\partial (\gamma_2-\gamma_4)}{\partial T} \right)s\frac{\partial T}{\partial s}+\gamma_2-2 \frac{\vartheta_2}{T}~.
\end{split}
\end{equation}
In the case of no external forces $\pi_i=0$, no parity odd terms and the bulk pressure at the surface 
vanishing $P|_{f=0}=0$, these constraints reduce to those in \cite{Armas:2013goa}, 
while, instead, if we have parity odd terms and require no bending corrections $\vartheta_1=\vartheta_2=0$, 
hence only $\alpha_3,\gamma_3,\gamma_6$ remain, 
then this reduces to the result of \cite{Jensen:2011xb, Jensen:2012jh}. 
There are a total of 10 relations in \eqref{eq:relations} relating the 6 surface transport coefficients $\alpha_1,\alpha_2,\alpha_3,\alpha_4,\vartheta_1,\vartheta_2$ to the 7 external coefficients $\pi_i$ and the 6 entropy current coefficients $\gamma_i$. Of these relations, 3 of them recover relations between bulk transport coefficients which were already known from a bulk analysis \cite{Bhattacharyya:2012nq}. Specifically, these are the second relation in the first line, the second relation in the second line and the first relation in the third line of \eqref{eq:relations}. From the 7 remaining relations, 6 of them determine the transport coefficients $\alpha_1,\alpha_3,\alpha_4$ in terms of the transport coefficients $\vartheta_1,\vartheta_2$ and the external coefficients $\pi_i$ while the remaining relation, namely the second relation in the third line of \eqref{eq:relations}, relates the transport coefficient $\alpha_3$ with the linear combination $\gamma_7$ defined in \eqref{eq:linear} in terms of the entropy current coefficients $\gamma_3,\gamma_6$ as observed in \cite{Jensen:2011xb, Jensen:2012jh}. Therefore, all surface coefficients appearing in \eqref{eq:Tab}-\eqref{eq:s_surf} are determined in terms of the 3 transport coefficients $\alpha_3,\vartheta_1,\vartheta_2$ and 3 independent $\pi_i$ coefficients, as expected, since it is indeed the number of independent scalars in both the 
partition function \eqref{surpartsecord} and the action \eqref{eq:action}.

Furthermore, from the partition function analysis, a total of 
28 constraints were obtained in \eqref{mainconst}. 
From the analysis of Appendix \ref{ssec:aaa}, we have obtained 
a total of 14 constraints from \eqref{eq:c1},\eqref{eq:c2} and \eqref{eq:c3}. 
However, as explained in Appendix \ref{ssec:aaa} the coefficients $\mathfrak{v1}_{7}$ and $\mathfrak{v2}_{7}$ may be removed by the transformation 
\eqref{Ttrans} and constitutes 2 of the 28 constraints in \eqref{mainconst}. 
The fact that \eqref{surPhenoST} is not in the Landau gauge adds 9 extra constraints. 
Therefore, we have that 28-14-2-9=3, which is exactly the number of constraints that we 
have obtained from an entropy current analysis.

\subsection{Comparison with the action and the partition function}
We now compare the constraints obtained in \eqref{eq:relations} with the results obtained from the action \eqref{eq:action} and the partition function \eqref{surpartsecord}. We begin by comparing \eqref{eq:genentropy} with the general form of the entropy current introduced in \cite{Bhattacharyya:2012nq}. Using the notation of \cite{Banerjee:2012iz}, the entropy current up to second order, and ignoring the first order corrections which vanish in equilibrium, can be written as
\begin{equation}
\begin{split}
J^{\mu}_{s(0)}=&~\tilde s \tilde u^{\mu} +\nabla_\nu\left(A_1\left(\tilde u^{\mu}\nabla^{\nu} \tilde T-\tilde u^{\nu}\nabla^{\mu}\tilde T\right)\right)+\nabla_\nu\left(A_2\tilde T \tilde{\omega}^{\mu\nu}\right)+A_3\left(\tilde R^{\mu\nu}-\frac{1}{2}g^{\mu\nu}\tilde R\right)\tilde u_\nu \\
&+\left(A_4\tilde u^{\nu}\nabla_\nu\tilde{\Theta} +A_5 \tilde R+A_6 \tilde R^{\mu\nu}\tilde u_{\mu}\tilde u_{\nu}+B_1\tilde{\omega}^2+B_2\tilde{\Theta}^2+B_3\tilde{\sigma}^2+B_4\nabla _\nu \tilde s\nabla^{\nu}\tilde s\right)\tilde u^{\mu} \\
&+2B_4\tilde s\tilde{\Theta}\nabla^{\mu}\tilde s+\left(\tilde{\Theta}\nabla^\mu B_5-P^{\lambda\rho}\nabla_\lambda \tilde u^{\mu}\nabla_\rho B_5\right)+B_6\tilde{\Theta} \tilde{\mathfrak{a}}^{\mu}+B_7\tilde{\mathfrak{a}}_\nu\tilde{\sigma}^{\mu\nu}~~\\
&+\tilde s~n^{\mu}\left( v_1 ~\tilde{\mathfrak{a}}_{\alpha} \tilde \omega^{\alpha \nu} n_{\nu}+  v_2  ~P^{\nu}_{\alpha} \nabla_{\beta} \tilde \omega^{\beta \alpha} n_{\nu} \right)~~,
\end{split}
\end{equation}
where the last line above is, again, due to the frame change \eqref{frmtransfull}. Since we are dealing with stationary configurations for which $\Theta=\sigma^{\mu\nu}=0$ we can ignore the terms involving $A_4,B_2,B_3,B_6,B_7$. We now contract this bulk entropy current with $n_\mu$ and evaluate it at the boundary $f=0$ imposing the boundary conditions \eqref{eq:boundaryc}. We find,
\begin{equation}
\begin{split}
J^{\mu}_{s(0)}n_\mu|_{f=0}=&~\left(A_1 T-\frac{\partial B_5}{\partial T} T-\left(A_1+\frac{\partial A_1}{\partial T}T+2TB_4\left(\frac{\partial \tilde s}{\partial \tilde T}\right)^{2} \right)s\frac{\partial T}{\partial s}\right)\Theta\mathfrak{a}^{\mu}n_\mu\\
&-T\left(A_2+\frac{\partial A_2}{\partial T}T\right)\tilde \omega^{\mu b}n_\mu\mathfrak{a}_b+A_1 Tu^{\mu}\nabla_\mu\left(\mathfrak{a}^{\mu}n_\mu\right)-\frac{\partial B_5}{\partial T} Tu^{a}\mathfrak{a}^{b}K_{ab}\\
&+A_2 T\nabla_a\left(\tilde \omega^{\mu a}n_\mu\right)-A_3\left(u^{b}\nabla_bK-u^{b}\nabla_a {K^{a}}_b\right)~~\\
&+\tilde s\left( v_1 ~\tilde{\mathfrak{a}}_{\alpha} \tilde \omega^{\alpha \nu} n_{\nu}+  v_2  ~P^{\nu}_{\alpha} \nabla_{\beta} \tilde \omega^{\beta \alpha} n_{\nu} \right)~~.
\end{split}
\end{equation}
Comparison of this with \eqref{eq:genentropy} we find read off
\begin{equation} \label{eq:id1}
\begin{split}
&\pi_1=A_1 T-\frac{\partial B_5}{\partial T} T-(A_1+\frac{\partial A_1}{\partial T}T+2TB_4\left(\frac{\partial \tilde s}{\partial \tilde T}\right)^{2})s\frac{\partial T}{\partial s}~~,~~\pi_2=-T\left(A_2+\frac{\partial A_2}{\partial T}T\right)~~, \\
&\pi_3=A_1 T~~,~~\pi_4=-\frac{\partial B_5}{\partial T} T~~,~~\pi_5=A_2 T~~,~~\pi_6=-A_3~~,~~\pi_7=A_3~~.
\end{split}
\end{equation}
By obtaining the bulk stress tensor and entropy current from the action \eqref{eq:action} and going to the orthogonal-Landau frame we obtain\footnote{Note that this identification is different than the one in \cite{Banerjee:2012iz}. As explained below \eqref{eq:newc}, this is because we have added divergence free terms to the bulk entropy current obtained from the action to set it in the form $J^{\mu}_{s(0)}\propto \tilde u^{\mu}$.}
\begin{equation}\label{eq:id2}
\begin{split}
&A_1 T=-2\frac{\partial \tilde P_1}{\partial T}+2\tilde P_3~~,~~A_2 T=-\frac{2\tilde P_2}{T}~~,~~A_3 T=-2{\tilde P_1}~~, \\
&B_4 T=-\frac{2}{T}\frac{\partial \tilde P_1}{\partial T}\left(\frac{\partial \tilde T}{\partial \tilde s}\right)^{2}~~,~\frac{\partial B_5}{\partial T}=-\frac{2}{T}\frac{\partial \tilde P_1}{\partial T}~~.
\end{split}
\end{equation}
Moreover, obtaining the surface stress tensor and surface entropy current from \eqref{eq:action}, setting it in the Landau frame and comparing it with \eqref{eq:Tab} and \eqref{eq:s_surf} leads to
\begin{equation}\label{eq:id3}
\begin{split}
&\gamma_1 T=-\tilde{\mathcal{B}}_3~~,~~\gamma_2 T=2\tilde P_3-\tilde{\mathcal{B}}_1-2T \frac{\partial \tilde P_1}{\partial T}~,~\gamma_3 T=2\tilde{\mathcal{B}}_2~~,~~\gamma_4 T=-2(\tilde P_1-\tilde{\mathcal{B}}_3)~~,~~\gamma_5 T=-2\tilde P_2~~,\\
&\gamma_7=2\tilde{\mathcal{B}}_2-\frac{\partial \tilde{\mathcal{B}}_2}{ \partial T} T~~, \vartheta_1=2\tilde P_1-\tilde{\mathcal{B}}_3~~,~~\vartheta_2=\tilde{\mathcal{B}}_1~~,~~\alpha_1=-\tilde{\mathcal{B}}_3+\frac{s}{T}\frac{\partial T}{\partial s}(-\tilde{\mathcal{B}}_3+\frac{\partial \tilde{\mathcal{B}}_3}{\partial T}T)~~,\\
&\alpha_2=(\tilde{\mathcal{B}}_1-2T\frac{\partial \tilde P_1}{\partial T})+\frac{s}{T}\frac{\partial T}{\partial s}\left(2\tilde P_3-2\frac{\partial \tilde P_1}{\partial T}T-\frac{\partial \tilde{\mathcal{B}}_1}{\partial T} T-\tilde{\mathcal{B}}_1+2(\tilde P_1-\tilde{\mathcal{B}}_3)\right)~~,\\
&\alpha_3=\frac{s}{T}\frac{\partial T}{\partial s}\left(2\tilde{\mathcal{B}}_2-\frac{\partial \tilde{\mathcal{B}}_2}{\partial T} T\right)~~.
\end{split}
\end{equation}
One can easily check that \eqref{eq:id1}-\eqref{eq:id3} satisfy the constraints \eqref{eq:relations}. We now turn into the partition function analysis of Sec.~\ref{ssec:frstordpartrel} and recast the relations \eqref{eq:relations} in terms of the transport coefficients written in \eqref{surPhenoST}. First, taking the stress tensor \eqref{surPhenoST}, we compute \eqref{eq:Tab} using \eqref{eq:Tabc} and then set it in the Landau frame. We find the stress tensor
\begin{equation} \label{eq:id21}
\begin{split}
\mathbb{T}^{ab}_{\text{sur}} =&~\mathcal{P}^{ab}\left( \left(\tilde{\mathfrak{s1}}_1+\frac{\tilde{\mathfrak{t}}}{2}-\frac{s}{T}\frac{\partial T}{\partial s}\tilde{\mathfrak{s2}}_1\right)\mathfrak{a}^{\mu}n_\mu +\left({\mathfrak{s1}}_2-\frac{\tilde{\mathfrak{t}}}{2}-\frac{s}{T}\frac{\partial T}{\partial s}\tilde{\mathfrak{s2}}_2\right)K+\left({\mathfrak{s1}}_3-\frac{s}{T}\frac{\partial T}{\partial s}{\mathfrak{s2}}_3\right)\ell^{\mu}n_\mu\right)\\
&+\tilde{\mathfrak{t}}~\mathcal{P}^{a}_c\mathcal{P}^{b}_cK^{cd}~~, 
\end{split}
\end{equation}
where we have defined
\begin{equation}\label{eq:id22}
\tilde{\mathfrak{s1}}_1={\mathfrak{s1}}_1-\frac{\partial \mathfrak{s1}_4}{\partial T} T-{\mathfrak{s1}}_4~~,~~\tilde{\mathfrak{s2}}_1={\mathfrak{s2}}_1+\frac{\partial \mathfrak{s2}_4}{\partial T} T+{\mathfrak{s1}}_4+2{\mathfrak{s2}}_4~~,~~\tilde{\mathfrak{s2}}_2={\mathfrak{s2}}_2-{\mathfrak{s2}}_4~~,~~\tilde{\mathfrak{t}}={\mathfrak{t}}+2{\mathfrak{s1}}_4~~.
\end{equation}
From here, upon comparison with \eqref{eq:Tab} we read off
\begin{equation}
\alpha_1=\tilde{\mathfrak{s1}}_1+\frac{\tilde{\mathfrak{t}}}{2}-\frac{s}{T}\frac{\partial T}{\partial s}\tilde{\mathfrak{s2}}_1~~,~~\alpha_2={\mathfrak{s1}}_2-\frac{\tilde{\mathfrak{t}}}{2}-\frac{s}{T}\frac{\partial T}{\partial s}\tilde{\mathfrak{s2}}_2~~,~~\alpha_3={\mathfrak{s1}}_3-\frac{s}{T}\frac{\partial T}{\partial s}{\mathfrak{s2}}_3~~,~~\alpha_4=\tilde{\mathfrak{t}}~~.
\end{equation}
Furthermore, from \eqref{eq:gstress2} we read off the components of $T^{ab\rho}_{(2)}$, 
\begin{equation}\label{eq:id23}
\vartheta_1=\mathfrak{s1}_4~~,~~\vartheta_2=\mathfrak{s1}_4+\mathfrak{s2}_4~~.
\end{equation}
Again, we can check that \eqref{eq:id21}-\eqref{eq:id23} satisfy the constraints \eqref{eq:relations}.

% %*******************************************************************************
 \section{Few useful relations}\label{app:NotandRel}
% %*******************************************************************************
%
Under a time independent diffeomorphism $x^{\mu} \rightarrow x^{\mu}+\epsilon^{\mu}(x^i)$ the background metric $\mathcal{G}_{\mu\nu}$ and the gauge field $\mathcal{A}_{\mu}$ transform as
\begin{equation}
 \delta \mathcal G_{\mu \nu} = -(\nabla_{\mu}\epsilon_{\nu}+\nabla_{\nu}\epsilon_{\mu}), ~\delta \mathcal A_{\mu} = -(\nabla_{\mu}\epsilon^{\nu}\mathcal A_{\nu}+\epsilon^{\nu}\nabla_{\nu}\mathcal A_{\mu})~~.
\end{equation}
As was noted earlier an equivalent description is to consider the background fields as $\sigma,~ a_i, ~g_{ij} $ for the metric and $A_0,~A_i$ for the gauge field.  Here we list out how these background fields transform under the time independent diffeomorphism written above
\begin{equation}
\begin{split}
\delta\sigma = -\epsilon^i \partial_i \sigma ~~,
& \delta a_i = -\partial_i \epsilon^0 -a_k \partial_i \epsilon^k -\epsilon^k \partial_k a_i~~, 
\delta g^{ij} = \nabla^i \epsilon^j + \nabla^j \epsilon^i ~~,\\
\delta A_0 &= -\epsilon^i \partial_i A_0 ~~,
\delta A_i = -\epsilon^k \nabla_k A_i -\nabla_i \epsilon^k A_k~~.
\end{split}
\end{equation}

%%%%%%%%%%%%%%%%%%%%%%%%%%%%

The fluid configurations discussed in section \ref{sec:config} were in 2+1 dimensional flat space spanned by 
coordinates $(t,~r,~\phi)$ with a metric 
\begin{equation} \label{configmet1}
ds^2 = -dt^2+dr^2+r^2d\phi^2~~.
\end{equation}
We would like to change coordinates to $(t,~r,~\tilde{\phi})$, such that $t,~r$ remains unchanged but
\begin{equation}
\tilde{\phi}= \phi - \omega t~~.
\end{equation}
In terms of coordinates $(t,~r,~\tilde{\phi})$, the metric in \eqref{configmet1} becomes
\begin{equation}
ds^2 = -{1\over \gamma^2} (dt - r^2 \omega \gamma^2 d\tilde{\phi})^2+ dr^2+r^2 \gamma^2 d\tilde{\phi}^2~~.
\end{equation}
Now comparing with \eqref{bkmet} we obtain the corresponding background fields as given below
\begin{equation}
\begin{split}
e^{2 \sigma} = {1 \over \gamma^2}&~~, ~~ a_{\tilde{\phi}}= -r^2 \omega \gamma^2~~,~~ a_r =0 ~~,
g_{\phi \phi} = r^2 \gamma^2~~,~~ g_{rr}=1~~.
\end{split}
\end{equation}

%%%%%%%%%%%%%%%%%%%%%%%%%%%%%

%%%%%%%%%%%%%%%%%%%%%%%%%%%%%
%\bibliographystyle{JHEP}
%\bibliography{SurfaceHydro}
\providecommand{\href}[2]{#2}\begingroup\raggedright\endgroup

\end{document}